\documentclass[a4paper]{article}
\usepackage[a4paper,top=3cm,bottom=2cm,left=3cm,right=3cm,marginparwidth=1.75cm]{geometry}
\usepackage{amsfonts}
\usepackage{bm}
\usepackage{extarrows}
\usepackage{amssymb}
\usepackage{color}
\usepackage[all]{xy}
\usepackage{graphicx}
\usepackage{braket}
\usepackage{amsmath}
\usepackage{appendix}
\usepackage{graphicx}
\usepackage[colorinlistoftodos]{todonotes}
\usepackage{amsthm}
\usepackage{mathrsfs}
\usepackage{amssymb}
\usepackage{accents}
\usepackage{extarrows}
\usepackage{makecell}
\usepackage{authblk}
\usepackage{url}
\usepackage{graphicx}
\usepackage{hyperref}
\usepackage{cite}
\usepackage{eufrak}
\hypersetup{colorlinks,
            citecolor=black,
            linkcolor=black,
            urlcolor=green,
            pdftex}

\def\be{\begin{equation}}
\def\ee{\end{equation}}
\def\ba{\begin{eqnarray}}
\def\ea{\end{eqnarray}}

\usepackage[english]{babel}
\usepackage[utf8x]{inputenc}
\usepackage[T1]{fontenc}

\usepackage{tikz}
\usetikzlibrary{positioning, shapes.geometric}

\usepackage[normalem]{ulem}

\title{The weak coupling theory of all dimensional loop quantum gravity}
\author[1,2]{Gaoping Long \footnote{201731140005@mail.bnu.edu.cn}}
\author[2]{Chun-Yen Lin \footnote{cynlin@ucdavis.edu}\thanks{corresponding author}}

\affil[1]{Department of Physics, South China University of Technology, Guangzhou 510641, China}
\affil[2]{Department of Physics, Beijing Normal University, Beijing 100875, China}

\date{}

\begin{document}

\maketitle

\begin{abstract}
The weak coupling loop quantum theory with Abelian gauge group provides us a new perspective to
study the weak coupling properties of LQG. In this paper, the weak coupling theory of all dimensional loop quantum gravity is established based on a symplectic-morphism between the $SO(D+1)$ holonomy-flux phase space and the $U(1)^{\frac{D(D+1)}{2}}$ holonomy-flux phase space. More explicitly, the Gaussian, simplicity, diffeomorphism and scalar constraint operators in $SO(D+1)$ loop quantum gravity will be generalized to the $U(1)^{\frac{D(D+1)}{2}}$ loop quantum theory based on the symplectic-morphism, and the $U(1)^{\frac{D(D+1)}{2}}$ loop quantum theory equipped with these constraint operators gives the weak coupling $U(1)^{\frac{D(D+1)}{2}}$ loop quantum gravity, with the corresponding Hilbert space is composed by the $U(1)^{\frac{D(D+1)}{2}}$ heat-kernel coherent states which are peaked at the weak coupling region of the $U(1)^{\frac{D(D+1)}{2}}$ holonomy-flux phase space.
\end{abstract}

\section{Introduction}
Loop quantum gravity (LQG) opens a convincing approach to achieve the unification of general relativity (GR) and quantum mechanics \cite{Ashtekar2012Background,RovelliBook2,Han2005FUNDAMENTAL,thiemann2007modern,rovelli2007quantum}. The distinguished feature of LQG is its non-perturbative and background-independent construction, which predicts the discretization of spatial geometry. An interesting research topic in the field is the weak coupling limit LQG, which is given by taking the limit that the Newton's gravitational constant $\kappa$ tends to 0. This idea was firstly proposed by Smolin and further studied by Tomlin and Varadarian \cite{Smolin_1992,PhysRevD.87.044039}. The resulting weak coupling LQG is a $U(1)^3$ gauge theory instead of the original $SU(2)$ gauge theory. This $U(1)^3$ LQG theory inherits some of the core characters of the original $SU(2)$ LQG, such as the discrete spatial geometry and the polymer-like quantization scheme. It has been used as a toy model to study the faithful LQG-like representation of the constraint algebra in the weak coupling limit of Euclidean GR \cite{Lewandowski:2016lby}. The theoretical framework of the weak coupling $U(1)^3$ LQG model is also used to study the quantum field theory on curved spacetime limit of LQG \cite{Sahlmann:2002qj,Sahlmann:2002qk}. Besides, it has been verified that the effective dynamics based on coherent state path-integral of the $U(1)^3$ LQG are consistent with that of the $SU(2)$ LQG in the weak coupling and semi-classical limit, with the Hamiltonian operators being defined accordingly \cite{Long:2021izw}.

The $SU(2)$ LQG only describes the quantum theory of GR in four dimensional spacetime. Nevertheless, our interests are beyond the quantum gravity in four dimensional spacetime, since various studies of classical and quantum gravity theories in higher-dimensional spacetimes (e.g., Kaluza-Klein theory, supergravity and superstring theories) show remarkable potentials in unifying the gravity and matter fields at the energy scale of quantum gravity. Thus, upon the background-independent and non-perturbative construction of the spacetime quantum geometry, it is an interesting task to extend the loop quantum method to the gravity theory in higher-dimensional spacetime, to explore a novel approach toward the ideas of unification. Pioneered by Bodendorfer, Thiemann and Thurn, the basic framework of loop quantization of GR in all dimensions has been established \cite{Bodendorfer:Ha,Bodendorfer:La,Bodendorfer:Qu,Bodendorfer:SgI}.
The resulting $(1+D)$-dimensional $SO(D+1)$ LQG takes the similar framework as
the standard (1+3)-dimensional $SU(2)$ LQG, i.e. the formulation of Yang-Mills gauge theory and the
loop quantization strategy. The most important difference between these two loop quantum theories is that the $(1+D)$-dimensional LQG contains an additional simplicity constraint, while the standard (1+3)-dimensional $SU(2)$ LQG only contains the Gaussian, and the ADM constraints.
It is well-known that the study of standard (1+3)-dimensional $SU(2)$ LQG encounters the difficulties of the quantum anomaly of the ADM constraints. However, the appearance of simplicity constraint in all dimensional LQG leads that
 the challenge of loop quantum anomaly already exists at the kinematic level before the accounts of the quantum ADM constraints. More explicitly, the classical connection phase space of all dimensional LQG is coordinatized by the canonical pairs $(A_{aIJ},\pi^{bKL})$, consisting of the spatial $so(D+1)$ valued connection fields $A_{aIJ}$ and the vector fields $\pi^{bKL}$. In the connection formulation, this theory is governed by a first class constraint system composed by the $SO(D+1)$ Gaussian constraint, the ADM constraints of $(1+D)$-dimensional GR, and an additional constraint called the simplicity constraint which takes the form $S^{ab}_{IJKL}:=\pi^{a[IJ}\pi^{|b|KL]}$ and generates extra gauge transformation in the $SO(D+1)$ connection phase space \cite{Bodendorfer:Ha,Bodendorfer:Qu}. Especially, it has been verified that the symplectic reductions with respected to the Gaussian and simplicity constraints in the $SO(D+1)$ connection  phase space lead to the familiar ADM phase space of  $(1+D)$-dimensional GR. Following similar procedures as that of the $SU(2)$ LQG, the loop quantization of the $SO(D+1)$ connection  formulation of all dimensional LQG gives the Hilbert space composed by the spin-network states of the $SO(D+1)$ holonomies, with the quantum numbers labeling these states carry the quanta of the flux operators representing the flux of $\pi^{bKL}$ over $(D-1)$-dimensional faces. Accordingly, the Gaussian and simplicity constraints can be promoted as operators acting in this Hilbert space.  
 Though the simplicity constraint is well-behaved in the classical connection formulation, it brings new challenges in the quantum gauge reduction procedures--- the quantum algebra among simplicity constraint operators in all dimensional LQG carries serious quantum anomaly. Specifically, the commutative Poisson algebra among the classical simplicity constraints becomes the deformed quantum algebra among the quantum simplicity constraint which is not even close \cite{Bodendorfer:2011onthe}. Moreover, it has been verified that the ``gauge'' transformations generated by these anomalous quantum simplicity constraints could connect the quantum states which are supposed to be physically distinct in terms of the semiclassical limit. Thus, the strong imposition of the anomalous quantum simplicity constraint leads to over-constrained quantum state space, which are not able to represent correct quantum geometry interpretation of this theory.

More explicit studies show that, the regularization and quantization of the classical simplicity constraints give two sets of local quantum simplicity constraints in all dimensional LQG, including the edge-simplicity constraint and the vertex-simplicity constraint. The anomaly of quantum algebra only appears for the vertex-simplicity constraint, while the edge-simplicity constraint remains anomaly free in the sense of taking a weakly commutative quantum algebra. The quantum anomaly of the vertex simplicity constraint can be revealed in the discrete phase space coordinatized by $SO(D+1)$ holonomy-flux variables faithfully. In other words, the Poisson algebras of simplicity constraint are isomorphic to quantum algebras of simplicity constraint, thus the anomaly of vertex-simplicity constraint already exists in the classical holonomy-flux phase space. Based on the so-called generalized twisted geometric parametrization of the edge-simplicity constraint surface, the gauge reduction with respect to  the simplicity constraint can be proceeded in the holonomy-flux phase space \cite{PhysRevD.103.086016}.
The result shows that, the discretized classical Gaussian, edge-simplicity constraints and vertex-simplicity
constraint which catches the anomaly of quantum vertex simplicity constraint define a constraint surface
in the holonomy-flux phase space of all dimensional LQG, and the kinematical physical degrees of freedom
 are captured by the gauge orbits in
the constraint surface generated by the first class system consisting of discretized Gaussian and edge-simplicity
constraints. Moreover, with the dual network partitioning the $D$-hypersurface, the reduced twisted geometry describes the geometric information of the dual network, which includes the $(D-1)$-faces' areas, the shape of each single $D$-polytope and the extrinsic curvature between arbitrary two adjacent  $D$-polytopes. Finally, the discrete ADM data
of the $D$-hypersurface in the form of Regge geometry can be identified as the degrees of freedom of the reduced generalized
twisted geometry space, up to an additional condition called the shape matching condition of $(D-1)$-dimensional faces.  Following this result, this gauge reduction procedures can be realized in quantum theory by imposing the quantum Gaussian and edge-simplicity constraint strongly, and imposing the vertex-simplicity constraint weakly. It leads to the physical kinematic Hilbert space spanned by the spin-network states labelled by simple representations at edges and gauge invariant simple coherent intertwiners at vertices \cite{long2019coherent}.
However, this treatment of the quantum gauge reduction with respect to quantum simplicity constraint introduces another issue.
%Usually, by proceeding the regularization and quantization procedures in LQG, a gauge invariant quantity in connection phase space can be promoted as an operator acting in the physical kinematic Hilbert space, and one could expect that the gauge degrees of freedom are identically eliminated in both classical and quantum case. However, it is not the case for simplicity constraint,
Notice that the gauge degrees of freedom with respect to simplicity constraint are eliminated by gauge fixing in classical connection theory, while they are eliminated by taking averaging with respect to gauge transformations in quantum theory.  Though the edge-simplicity constraints only transform the pure-gauge components in the holonomy, the gauge reduction by taking gauge averaging destroys the structure of holonomy, which leads that the simplicity reduced holonomy can not capture the degrees of freedom of intrinsic curvature. In other words, the simplicity reduced holonomy is not able to inherit the property of connection and thus it can not be used as the building block to regularize the connection.

 In principle, this problem of the simplicity reduced holonomy can be tackled in two strategies. In the first strategy, one can re-construct a gauge invariant holonomy with respect to the simplicity constraint to ensure that it captures the the degrees of freedom of intrinsic and extrinsic curvature properly, by following the geometric interpretation of each component of holonomy given by the twisted geometry parametrization \cite{PhysRevD.103.086016}. More explicitly, in order to ensure that the gauge invariant holonomy with respect to the simplicity constraint is able to capture the degrees of freedom of intrinsic curvature, one need to add some terms involving the holonomy of Levi-Civita connection to the simplicity reduced holonomy. This strategy has been considered in our previous work \cite{Long:2022thb}, and we find that the operator corresponds to Levi-Civita connection would be a rather complicated function of flux operator, thus it still need further researches. The second strategy is to proceed the quantum gauge reduction with respect to simplicity constraint by using the gauge fixing scheme, so that the gauge degrees of freedom are eliminated consistently in both connection theory and quantum theory. Usually, the gauge fixing scheme could be proceeded at the semi-classical level  based on the coherent states whose wave functions converge along the gauge orbits of simplicity constraint sharply. However, such kind of coherent states in $SO(D+1)$ LQG must involve the non-simple representations of $SO(D+1)$, which leads that the gauge fixing scheme encounter intractable technical difficulties.

\begin{figure}
\centering
\tikzstyle{startstop} = [rectangle, rounded corners, minimum width=3cm, minimum height=1cm,text centered, draw=black]
\tikzstyle{decision} = [diamond, draw, text width=5.5em, text badly centered, inner sep=0pt]
\tikzstyle{process} = [rectangle, minimum width=3cm, minimum height=1cm, text centered, draw=black]
\tikzstyle{arrow} = [thick,->,>=stealth]
\begin{tikzpicture}[node distance = 1.5cm]
% Place nodes
\node (start) [startstop, align=center] {Connection formulation of \\ $(1+D)$-dimensional GR};
\node (LQ) [decision, below of = start, yshift=-1.0cm] {Loop representation};
\node (HF1) [process, left of=LQ, xshift=-3.0cm, yshift=-2cm, align=center] { $SO(D+1)$ holonomy \\ -flux  phase space};
\node (HF2) [process, right of=LQ, xshift=3.0cm, yshift=-2cm, align=center] {$U(1)^{\frac{D(D+1)}{2}}$ holonomy \\ -flux  phase space};

%\node (decLambda) [decision, below of =HF1, yshift=-0.5cm] {};
%\node (decAlpha) [decision, below of =HF2, yshift=-0.5cm] {$SO(D+1)$ Hilbert \\ space};

\node (HB1) [process, below of=HF1, yshift=-1.0cm, align=center] {$SO(D+1)$ Hilbert \\ space and \\ constraint operators};
\node (HB2) [process, below of=HF2, yshift=-1.0cm, align=center] {$U(1)^{\frac{D(D+1)}{2}}$ \\ Hilbert space};
\node (OP) [process, below of=HB1, yshift=-1.0cm, align=center] {Constraint operators \\ based on re-parametrization};
\node (HB3) [process, below of=HB2, yshift=-1.0cm, align=center] {The weak coupling \\ Hilbert sub-space};
\node (end) [startstop, below of=start, yshift=-10cm, align=center] {The weak coupling \\ $U(1)^{\frac{D(D+1)}{2}}$ LQG};
%\node (end) [startstop, below of=decLambda, xshift=3.5cm, yshift=-5.5cm] {Optimal};

% Draw edges
\draw [arrow] (start) -- (LQ);
\draw [arrow] (LQ) -| node[near start, above] {$so(D+1)$ basis} (HF1);
\draw [arrow] (LQ) -| node[near start, above] {$u(1)^{\frac{D(D+1)}{2}}$ basis} (HF2);
\draw [<->] (HF2) --node[anchor = north] {(i) Re-parametrization} (HF1);
%\draw[-] (n0.south) ¨C (n1); ²»´ø¼ýͷʵÏß
\draw [arrow](HF1) --node[anchor = east] {Quantization } (HB1);
\draw [arrow](HF2) --node[anchor = east] {Quantization } (HB2);
\draw [arrow](HB1) --node[anchor = east] {(ii)} (OP);
\draw [arrow](HB2) --node[anchor = east] {(iii)} (HB3);
%\draw [-] (decLambda) -- (DeleteConst);
%\draw [arrow] (HB3) -- +(-2.5,0) |- node[near start, left] {yes} (end);
%\draw [arrow] (decAlpha) -- node[anchor = east] {no} (AddConst);
\draw [arrow] (OP) --+(0,-0.5)|- (end);
\draw [arrow] (HB3) --+(0,-0.5)|-  (end);

%\draw [arrow](DeleteConst) |- (computeP);
%\draw [arrow](AddConst) |- (computeP);
%\draw [arrow](computeP) -- +(0,-1.5) -- +(7,-1.5) |- (0,-0.8);
\draw [arrow]+(0,-5.0) |- (-4.4,-8.4);
\end{tikzpicture}
\caption{Flow chart of the establishment of the weak coupling $U(1)^{\frac{D(D+1)}{2}}$ LQG. In step (i), the $U(1)^{\frac{D(D+1)}{2}}$ holonomy-flux variables provides a re-parametrization of the $SO(D+1)$ holonomy-flux phase space, and the $SO(D+1)$ holonomy-flux Poisson algebra can be re-produced by the re-parametrization based on the $U(1)^{\frac{D(D+1)}{2}}$ holonomy-flux Poisson algebra in the weak coupling limit. In step (ii), the constraint operators in $SO(D+1)$ LQG are generalized to the $U(1)^{\frac{D(D+1)}{2}}$ loop quantum theory based on the re-parametrization. In step (iii), the weak coupling Hilbert sub-space are composed by the coherent states in $U(1)^{\frac{D(D+1)}{2}}$ Hilbert space which are peaked at the weak coupling region of the $U(1)^{\frac{D(D+1)}{2}}$ holonomy-flux phase space.}
\label{fig: Flow chart}
\end{figure}
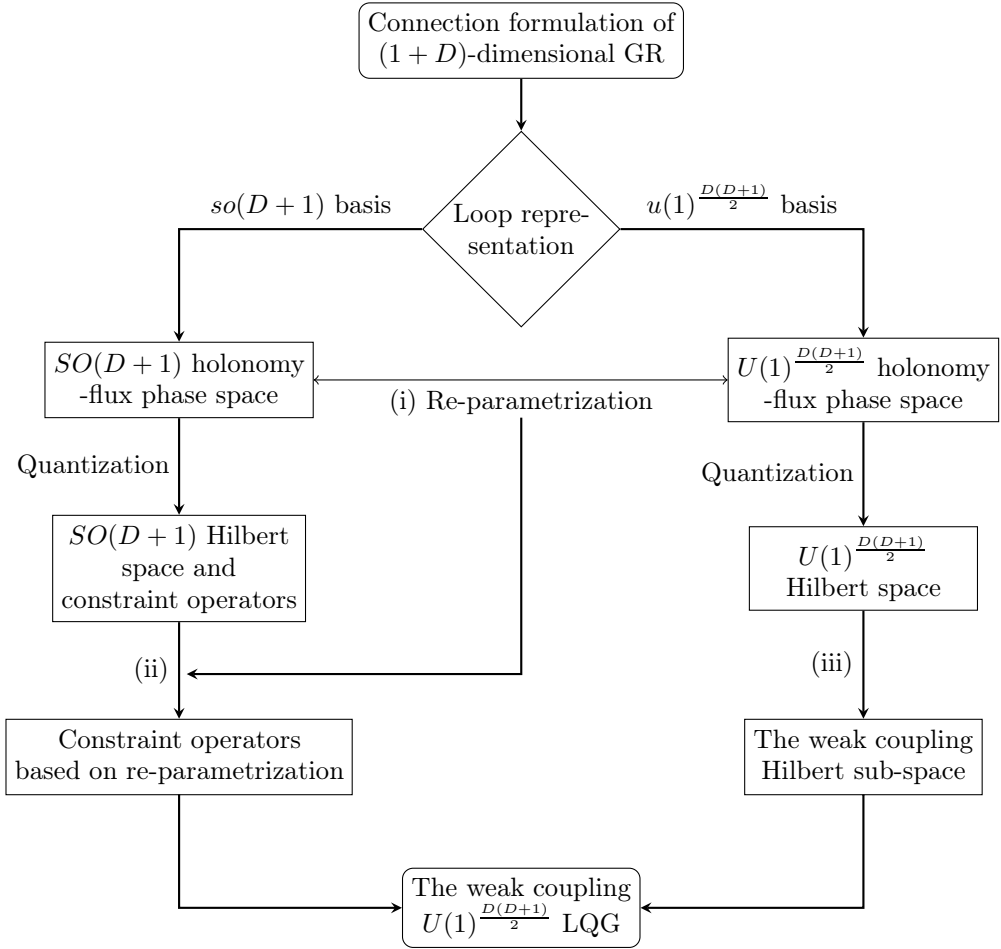

 The weak coupling theory of LQG equipped with Abelian gauge group  provides us a new perspective  to proceed the quantum gauge reduction with respect to simplicity constraint based on the gauge fixing scheme. Notice that the coherent states in the loop quantum theory equipped with Abelian gauge group is just a simple combination of the heat-kernel coherent state of $U(1)$. In this paper, we will show that the weak coupling theory of $SO(D+1)$ LQG can be reformulated as a loop quantum theory equipped with Abelian gauge group, so that one can avoid the obstacle introduced by the non-simple representations of $SO(D+1)$ and the gauge fixing with respect to simplicity constraint become feasible based on the heat-kernel coherent state of $U(1)$. More explicitly, we will consider the loop representation of the quantization of the connection formulation of $(1+D)$-dimensional GR ($D\geq3$), with the corresponding quantum algebra being given by the $U(1)^{\frac{D(D+1)}{2}}$ holonomy-flux variables. Since the Gaussian constraint in the connection formulation generates $SO(D+1)$ gauge transformations, the loop representation with $U(1)^{\frac{D(D+1)}{2}}$ holonomy-flux leads that the constraint operators are hardly to be defined. Nevertheless, this issue can be avoided in the weak coupling limit that the holonomies tend to identity. We will show that the $U(1)^{\frac{D(D+1)}{2}}$ holonomy-flux variables give a re-parametrization of the $SO(D+1)$ holonomy-flux phase space, and the $SO(D+1)$ holonomy-flux Poisson algebra can be re-produced by $U(1)^{\frac{D(D+1)}{2}}$ holonomy-flux Poisson algebra based on this re-parametrization in the weak coupling limit. Thus, the $U(1)^{\frac{D(D+1)}{2}}$ loop quantum theory can be regarded as another kind of quantization of the weak coupling region of $SO(D+1)$ holonomy-flux phase space in loop representation. Following this result, the Gaussian, simplicity, diffeomorphism and scalar constraint operators in $SO(D+1)$ LQG  will be generalized to the $U(1)^{\frac{D(D+1)}{2}}$ loop quantum theory based on the re-parametrization, and the $U(1)^{\frac{D(D+1)}{2}}$ loop quantum theory equipped with these constraint operators gives the weak coupling $U(1)^{\frac{D(D+1)}{2}}$ LQG, with the corresponding Hilbert space is composed by the $U(1)^{\frac{D(D+1)}{2}}$ heat-kernel coherent states which are peaked at the weak coupling region of the $U(1)^{\frac{D(D+1)}{2}}$ holonomy-flux phase space. These ideas are illustrated in Fig.\ref{fig: Flow chart}.

This paper is organized as follows. The elements of the classical theory and quantum theory of $SO(D+1)$ LQG will be introduced in section 2. Especially, we will emphasis the gauge reduction with respect to simplicity constraint and the issue in the construction of scalar constraint operator. Then in section 3, by introducing a privileged parametrization of the $SO(D+1)$ holonomy-flux phase space using the coordinates of the $U(1)^{\frac{D(D+1)}{2}}$ holonomy-flux phase space, and extending this parametrization as a symplectic-morphism, the weak coupling $U(1)^{\frac{D(D+1)}{2}}$ LQG will be constructed based on the Hilbert space composed by the $U(1)^{\frac{D(D+1)}{2}}$ heat-kernel coherent states peaked at the weak coupling region. Besides, we will discuss the treatment of the constraints in  the weak coupling $U(1)^{\frac{D(D+1)}{2}}$ LQG in section 4. Finally, we will finish with a conclusion and outlook in section 5.

\section{Elements of the $SO(D+1)$ LQG in (1+D)-dimensional spacetime}
\subsection{The connection phase space of $SO(D+1)$ LQG}\label{sec2.1}
The connection dynamics formulation of $(1+D)$-dimensional GR is given by the phase space coordinatized by the canonical field variables $(A_{aIJ}, \pi^{bKL})$ on a spatial $D$-dimensional manifold $\Sigma$, which is equipped with the kinematic constraints---Gaussian constraint $\mathcal{G}^{IJ}\approx0$ and simplicity constraint $S^{ab[IJKL]}\approx0$ inducing the gauge transformation of this theory, and the dynamics constraints---diffeomorphism constraint $C_a\approx0$ and scalar constraint $C\approx0$. The only non-trivial Poisson bracket between the conjugate pair reads \cite{Bodendorfer:Ha}
\begin{equation}\label{Poisson1}
\{{A}_{aIJ}(x), \pi^{bKL}(y)\}=2\kappa\gamma\delta_a^b\delta_{[I}^K\delta_{J]}^{L}\delta^{(D)}(x-y),
\end{equation}
where $\kappa$ is the Newton's gravitational constant, $\gamma$ is the Barbero-Immirzi parameter, and we introduced the notation $a,b,... = 1,2,...,D$ for the spatial tensorial indices and $I,J,... = 1,2,...,D + 1$
for the $so(D+1)$ Lie algebra indices in its definition representation.
 Based on the Poisson bracket \eqref{Poisson1}, one can check that the Gaussian constraint
\begin{equation}
\mathcal{G}^{IJ}:=\partial_a\pi^{aIJ}+2{A}_{aK}^{[I}\pi^{a|K|J]}\approx0,
\end{equation}
simplicity constraint
\begin{equation}
 S^{ab[IJKL]}:=\pi^{a[IJ}\pi^{|b|KL]}\approx0
\end{equation}
combining with the diffeomorphism constraint
 and scalar constraint
  form a first class constraint system in the connection phase space. It has been verified that the symplectic reduction with respect to the Gaussian and simplicity constraints in this connection phase space leads to the ADM phase space of  all dimensional GR. Indeed, as all of the Yang-Mills gauge theory, the Gaussian constraint generates the $SO(D+1)$ gauge transformation of the connection ${A}_{aIJ}$ and its momentum $\pi^{bKL}$. Specifically, the simplicity constraint restricts the degrees of freedom of $\pi^{aIJ}$ to that of a D-frame $E^{aI}$ describing the spatial internal geometry and generates some other gauge transformation. The connection variables can be related to the geometric variables on the constraint surface of both Gaussian and simplicity constraint. More explicitly, on the simplicity constraint surface one has $\pi^{aIJ}=2n^{[I}E^{|a|J]}$, with $E^{aI}$ being the densitized D-frame related to double densitized dual metric by $\tilde{\tilde{q}}^{ab}=E^{aI}E^b_I$, and $n^I$ being a unit internal vector defined by $n_IE^{aI}=0$. Besides, one can define the spin connection $\Gamma_{aIJ}$ as
\begin{equation}\label{spingamma}
\Gamma_{aIJ}[\pi]=\frac{2}{D-1}T_{aIJ}+ \frac{D-3}{D-1}\bar{T}_{aIJ}+\Gamma^b_{ac}T^c_{bIJ}
\end{equation}
  which satisfies $\partial_{a}e_{b}^{I}-\Gamma_{ab}^ce_c^I+\Gamma_{a}^{IJ}e_{bJ}=0$ on simplicity constraint surface,  where $T_{aIJ}:=\pi_{bK[I}\partial _a \pi^{bK}_{\ \ J]}$,  $T^c_{bIJ}:= \pi_{bK[I}\pi^{cK}_{\ \ J]}$, $\bar{T}_{aIJ}:=\bar{\eta}_I^K\bar{\eta}_J^LT_{aKL}$, $\bar{\eta}_I^J=\delta_I^J-n_In^J$, $\Gamma_{ab}^c$ is the Levi-Civita connection of $q_{ab}$ and $e_{aI}$ being the D-bein defined by $E^{aI}e_{bI}=\sqrt{q}\delta^a_b$. Based on these conventions, the densitized extrinsic curvature of the spatial manifold $\sigma$ can be given by
\begin{equation}
\tilde{{K}}_a^{\ b}={ K}_{aIJ}\pi^{bIJ}\equiv \frac{1}{\gamma}({A}_{aIJ}-\Gamma_{aIJ})\pi^{bIJ}
\end{equation}
 on the constraint surface of both Gaussian and simplicity constraint. Now, it is worth to clarify the gauge transformation induced by simplicity constraint. One can check that $A_{aIJ}$ transforms with respect to simplicity constraint as
\begin{eqnarray}\label{simgauge}
&&\int_{\sigma}d^Dx f_{ab[IJKL]}(x)\{S^{abIJKL}, {A}_{cMN}(y)\}\\\nonumber
&=&2\beta\kappa f_{ac[IJMN]}(y)\pi^{aIJ}(y)= 4\beta\kappa f_{ac[IJMN]}(y)n^{[I}E^{|a|J]}(y).
\end{eqnarray}
 on the simplicity constraint surface. In fact, by decomposing the connection ${A}_{aIJ}=2n_{[I}{A}_{|a|J]}+\bar{{A}}_{aIJ}$, it is easy to verify that, on the simplicity constraint surface, only the component $\bar{{A}}_{a}^{IJ}$ transforms while the component $2n_{[I}{A}_{|a|J]}$ is invariant under the transformation induced by simplicity constraint. Similarly, ${K}_{aIJ}
:=\frac{1}{\gamma}({A}_{aIJ}-\Gamma_{aIJ})$ can be decomposed as ${K}_{aIJ}=2n_{[I}{K}_{|a|J]}+\bar{{K}}_{aIJ}$. Also, one can verify that, on the simplicity constraint surface, the component $2n_{[I}{K}_{|a|J]}$ is gauge invariant while $\bar{{K}}_{a}^{IJ}$ is gauge variant with respect to simplicity constraint. Hence, the simplicity constraint fixes both $\tilde{{K}}_a^{\ b}$ and $q_{ab}$ so that it exactly introduce extra gauge degrees of freedom. Indeed, one can construct the simplicity reduced connection
\begin{equation}
A^S_{aIJ}:=A_{aIJ}-\gamma\bar{K}_{aIJ}
\end{equation}
which is gauge invariant with respect to simplicity constraint.
Then, the symplectic reduction with respect to the simplicity constraint in the connection phase space can be illustrated as
 $$\xymatrix{}
\xymatrix@C=3.5cm{
(A_{aIJ},\pi^{bKL})
\ar[r]^{\textrm{reduction}} &(A^S_{aIJ},\pi^{bKL})|_{S^{abIJKL}=0} },$$
which leads to the gauge invariant variables $(A^S_{aIJ},\pi^{bKL})$ with respect to simplicity constraint on the constraint surface defined by $S^{abIJKL}=0$. Here we would like to emphasis that the gauge reduction with respect to simplicity constraint is realized by taking gauge fixing, in other words, the pure gauge component $\bar{K}_{aIJ}$ is fixed by $\bar{K}_{aIJ}=0$.

Now let us turn to consider the explicit expression of the scalar constraint in the connection phase space.  Similar to the analogue in the connection  formulation of (1+3)-dimensional GR, one can establish the scalar constraint in the connection  formulation of $(1+D)$-dimensional GR based on two terms---the so called Euclidean term $C_{\text{E}}$ and Lorentzian term $C_{\text{L}}$ \cite{Bodendorfer:Qu}. The Euclidean term $C_{\text{E}}$ reads
\begin{equation}
C_{\text{E}}:=\frac{1}{\sqrt{\det(q)}}F_{abIJ}\pi^{aIK}{\pi^{b\ J}_K}
\end{equation}
with $
F_{abIJ}:=\partial_aA_{bIJ}-\partial_bA_{aIJ}+\delta^{KL}A_{aIK}A_{bLJ}-\delta^{KL}A_{aJK}A_{bLI}
$. Define
\begin{equation}
 C_{\text{E}}[1]:=\int_{\sigma}d^Dy C_{\text{E}}(y),
 \end{equation}
then the Lorentzian term $C_{\text{L}}$ reads
\begin{eqnarray}
C_{\text{L}}&:=&
-\frac{8(1+\gamma^2)}{\sqrt{\det(q)}}
K_{[a|I|}K_{b]J}E^{aI}E^{b J}\\\nonumber
&=&\frac{4(1+\gamma^2)}{\sqrt{\det(q)}}
[K_{b}^{\  a}K_{a}^{\ b}-K^2],
\end{eqnarray}
where $K(x):=K_{aI}(x)E^{aI}(x)$ and $K_{b}^{\  a}:=K_{bI}E^{aI}$ are given by
\begin{equation}
K(x)=-\frac{1}{4\kappa\gamma^2}\{C_{\text{E}}(x),V(x,\epsilon)\}
\end{equation}
and
\begin{equation}
K_{aI}(x)E^{bI}(x)=-\frac{1}{8\kappa^2\gamma^3}\pi^{bKL}(x)\{A_{aKL}(x), \{C_{\text{E}}[1],V(x,\epsilon)\}\}
\end{equation}
 on the constraint surface of both Gaussian and simplicity constraint,
 with $R(x,\epsilon)\ni x$ being a D-dimensional hyper-cube with coordinate scale $\epsilon$ and $V(x,\epsilon)$ being the volume of $R(x,\epsilon)$.
One can check that  $H_{\text{E}}$ contains the pure gauge component $\bar{K}_{aIJ}$ through the identity
\begin{eqnarray}
C_{\text{E}}&:=&\frac{1}{\sqrt{\det(q)}}F_{abIJ}\pi^{aIK}{\pi^{b\ J}_K}\\\nonumber
&=&-\sqrt{\det(q)}R-\frac{\gamma^2}{\sqrt{\det(q)}}(4[K_{b}^{\ a}K_a^{\ b}-K^2]+(\bar{K}_{bIK}E^{aI})(\bar{K}_{aJ}^{\ \ \, K}E^{bJ})),
\end{eqnarray}
which holds on the constraint surface of both Gaussian and simplicity constraint, where $R$ is the scalar curvature of $\Gamma_{aIJ}$ defined by
 \begin{equation}\label{scalarR}
   R:=-\frac{1}{\det(q)}R_{abIJ}\pi^{aIK}{\pi^{b\ J}_K}
 \end{equation}
 with $R_{abIJ}:=\partial_a\Gamma_{bIJ}-\partial_b\Gamma_{aIJ}+\delta^{KL}\Gamma_{aIK}\Gamma_{bLJ} -\delta^{KL}\Gamma_{aJK}\Gamma_{bLI}
$. Thus, in order to get the correct gauge invariant ADM scalar constraint on the constraint surface of both Gaussian and simplicity constraint,  the scalar constraint in $SO(D+1)$ connection  formulation of (1+D)-GR must contain an additional term $\frac{\gamma^2}{\sqrt{\det(q)}}(\bar{K}_{bIK}E^{aI})(\bar{K}_{aJ}^{\ \ \, K}E^{bJ})$  to cancel the gauge variant term in $C_{\text{E}}$.  The term $\frac{\gamma^2}{\sqrt{\det(q)}}(\bar{K}_{bIK}E^{aI})(\bar{K}_{aJ}^{\ \ \, K}E^{bJ})$ can also be expressed in connection variables by using that
\begin{equation}\label{KFD}
\bar{K}_{bKL}=\frac{4}{\gamma}(F^{-1})_{aIJ,bKL}\bar{D}^{aIJ}
\end{equation}
and
\begin{equation}
(\bar{K}_{bIK}E^{aI})(\bar{K}_{aJ}^{\ \ \, K}E^{bJ})=\frac{4}{\gamma^2}\bar{D}^{aIJ}(F^{-1})_{aIJ,bKL}\bar{D}^{bKL}
\end{equation}
holds on the simplicity constraint surface,
where we define
\begin{equation}\label{F-1}
(F^{-1})_{aIJ,bKL}:=\frac{1}{4(D-1)}\pi_{aAC}\pi_{bBD}(\pi^{cEC}\pi_{cE}^{\ \ D}-\delta^{CD})(\delta^{AB}\delta^{K[I}\delta^{J]L}-2\delta^{LA}\delta^{B[I}\delta^{J]K}),
\end{equation}
\begin{equation}\label{barDaIJ}
\bar{D}^{aIJ}:=\left(\delta^a_b\bar{\delta}^I_{[K}\bar{\delta}^J_{L]}+\frac{2}{D-1}(\pi^{aM[I} \bar{\delta}^{J]}_{[K} \pi_{bL]M}-\delta^a_b\delta^{[I}_{[K}n^{J]}n_{L]})\right)D^{bKL}
\end{equation}
and
\begin{equation}\label{DaIJ}
D^{aIJ}:=\pi^{b[I}_{\ \ \ K}\mathcal{D}_b\pi^{a|K|J]}
\end{equation}
with
\begin{equation}
\bar{\delta}^{J}_{K}= (\delta^J_K-n^Jn_K),\quad  n^In_J=\frac{1}{D-1}(\pi^{aKI}\pi_{aKJ}-\delta^I_J)
\end{equation}
 on the simplicity constraint surface. Moreover, $(F^{-1})_{aIJ,bKL}$ and $\bar{D}^{aIJ}$ can be extended as functionals on the entire connection phase space naturally.
Now, the final scalar constraint  reads
\begin{equation}\label{scalar1}
C=C_{\text{E}}+C_{\text{L}}+C_{\text{D}}
\end{equation}
with
\begin{equation}\label{CD}
C_{\text{D}}:=\frac{4}{\sqrt{\det(q)}}\bar{D}^{aIJ}(F^{-1})_{aIJ,bKL}\bar{D}^{bKL}.
\end{equation}

The scalar constraint also can be expressed in a simpler formulation by using the  simplicity reduced connection
\begin{equation}
A^S_{aIJ}\equiv A_{aIJ}-\gamma\bar{K}_{aIJ},
\end{equation}
whose curvature is defined by
\begin{equation}
F^S_{abIJ}:=\partial_aA^S_{bIJ}-\partial_bA^S_{aIJ}+\delta^{KL}A^S_{aIK}A^S_{bLJ} -\delta^{KL}A^S_{aJK}A^s_{bLI}.
\end{equation}
It is easy to check
\begin{equation}
C^S_{\text{E}}:=\frac{1}{\sqrt{\det(q)}}F^S_{abIJ}\pi^{aIK}{\pi^{b\ J}_K}=-\sqrt{\det(q)}R-\frac{4\gamma^2}{\sqrt{\det(q)}}[K_{ab}K^{ab}-K^2]
\end{equation}
and
\begin{equation}
K_{aI}(x)E^{bI}(x)=-\frac{1}{8\kappa^2\gamma^3}\pi^{bKL}(x)\{A_{aKL}(x), \{C^S_{\text{E}}[1],V(x,\epsilon)\}\}
\end{equation}
hold on the constraint surface of both Gaussian and simplicity constraint. Then, the scalar constraint can be expressed as
\begin{equation}\label{scalar2}
C=C^S_{\text{E}}+C_{\text{L}}.
\end{equation}
In fact, the $C_{\text{D}}$ term in \eqref{scalar1} offsets the gauge variant part in $C_{\text{E}}$ and it leads to $C^S_{\text{E}}=C_{\text{E}}+C_{\text{D}}$ exactly. The gauge degrees of freedom in the expressions \eqref{scalar1} and \eqref{scalar2} are eliminated by taking gauge fixing, which means, the gauge component $\bar{K}_{aIJ}$ with respect to simplicity constraint are fixed as zero. However, we will see that the gauge degrees of freedom will be eliminated by taking averaging with respect to the gauge transformation in quantum theory of $SO(D+1)$ LQG, which contradicts to the treatment for the scalar constraint and it introduces new obstacle to the construction of the scalar constraint operators.
\subsection{The discrete phase space of $SO(D+1)$ LQG}\label{sec2.2}
Since the gauge group $SO(D+1)$ of the connection formulation of $(1+D)$-dimensional GR is compact, the quantisation of the $SO(D+1)$ connection formulation is therefore in complete analogy with (1+3)-dimensional $SU(2)$ LQG \cite{Ashtekar2012Background,thiemann2007modern,rovelli2007quantum,RovelliBook2,Han2005FUNDAMENTAL}. Following arbitrary standard text on LQG such as \cite{thiemann2007modern,rovelli2007quantum}, the loop quantization of the $SO(D+1)$ connection formulation of $(1+D)$-dimensional GR gives the kinematical Hilbert space $\mathcal{H}$ \cite{Bodendorfer:Qu}. Indeed, this space can be regarded as a union of the Hilbert spaces $\mathcal{H}_\Gamma=L^2((SO(D+1))^{|E(\Gamma)|},d\mu_{\text{Haar}}^{|E(\Gamma)|})$ on all possible graphs $\Gamma$ embedded in $\Sigma$,  where $E(\Gamma)$ denotes the set composed by the independent edges of $\Gamma$ and $d\mu_{\text{Haar}}^{|E(\Gamma)|}$ denotes the product of the Haar measure on $SO(D+1)$. This result indicates that there is a discrete phase space $(T^\ast SO(D+1))^{|E(\Gamma)|}$ on each given $\Gamma$, which is coordinatized by the elementary discrete variables---holonomies and fluxes. The holonomy of $A_{aIJ}$ along an edge $e\in\Gamma$ is defined by
 \begin{equation}
h_e[A]:=\mathcal{P}\exp(\int_eA)=1+\sum_{n=1}^{\infty}\int_{0}^1dt_n\int_0^{t_n}dt_{n-1}...\int_0^{t_2} dt_1A(t_1)...A(t_n),
 \end{equation}
 where $A(t):=\frac{1}{2}\dot{e}^aA_{aIJ}\tau^{IJ}$, $\dot{e}^a$ is the tangent vector field of $e$, $\tau^{IJ}$ is a basis of $so(D+1)$ given by $(\tau^{IJ})^{\text{def.}}_{KL}=2\delta^{[I}_{K}\delta^{J]}_{L}$ in definition representation space of $SO(D+1)$, and $\mathcal{P}$ denoting the path-ordered product.
The flux $F^{IJ}_e$ of $\pi^{aIJ}$ through the $(D-1)$-dimensional face dual to edge $e$ in the perspective of source point of $e$ is defined by
\begin{equation}\label{F111}
 F^{IJ}_e:=-\frac{1}{4}\text{tr}\left(\tau^{IJ}\int_{e^\star}\epsilon_{aa_1...a_{D-1}}h(\rho^s_e(\sigma)) \pi^{aKL}(\sigma)\tau_{KL}h(\rho^s_e(\sigma)^{-1})\right),
 \end{equation}
 where $e^\star$ is the $(D-1)$-face traversed by $e$ in the dual lattice of $\Gamma$, $\rho_e^s(\sigma): [0,1]\rightarrow \Sigma$ is a path connecting the source point $s(e)\in e$ to $\sigma\in e^\star$ such that $\rho_e^s(\sigma): [0,\frac{1}{2}]\rightarrow e$ and $\rho_e^s(\sigma): [\frac{1}{2}, 1]\rightarrow e^\star$. Similarly, we can define the dimensionless flux $X^{IJ}_e$ as
 \begin{equation}\label{X}
 X^{IJ}_e=-\frac{1}{4\gamma a^{D-1}}\text{tr}\left(\tau^{IJ}\int_{e^\star}\epsilon_{aa_1...a_{D-1}}h(\rho^s_e(\sigma)) \pi^{aKL}(\sigma)\tau_{KL}h(\rho^s_e(\sigma)^{-1})\right),
 \end{equation}
 where $a$ is an arbitrary but fixed constant with the dimension of length.
  One can also define the dimensionless flux $\tilde{X}^{IJ}_e$ in the perspective of target point of $e$ as
  \begin{equation}\label{tildeX}
 \tilde{X}^{IJ}_e=\frac{1}{4\gamma a^{D-1}}\text{tr}\left(\tau^{IJ}\int_{e^\star}\epsilon_{aa_1...a_{D-1}}h(\rho^t_e(\sigma)) \pi^{aKL}(\sigma)\tau_{KL}h(\rho^t_e(\sigma)^{-1})\right),
 \end{equation}
 where $\rho_e^t(\sigma): [0,1]\rightarrow \Sigma$ is a path connecting the source point $t(e)\in e$ to $\sigma\in e^\star$ such that $\rho_e^t(\sigma): [0,\frac{1}{2}]\rightarrow e$ and $\rho_e^t(\sigma): [\frac{1}{2}, 1]\rightarrow e^\star$. It is easy to see that ${X}^{IJ}_e$ and $\tilde{X}^{IJ}_e$ have the relation 
 \begin{equation}\label{XXrelation}
   h_{e}^{-1}X^{IJ}_{e}\tau_{IJ}h_{e}=-\tilde{X}^{KL}_{e}\tau_{KL}.
 \end{equation}
 Notice that $SO(D+1)\times so(D+1)\cong T^\ast SO(D+1)$, the discrete phase space $\times_{e\in \Gamma}(SO(D+1)\times so(D+1))_e$ is a direct product of $SO(D+1)$ cotangent bundles, and it is referred to as the holonomy-flux phase space of $SO(D+1)$ loop quantum gravity on the fixed graph $\Gamma$. The complete phase space of the theory can be given by taking the union over the holonomy-flux phase spaces of all possible graphs.
In the discrete holonomy-flux phase space associated to $\Gamma$, the constraints are expressed by the holonomy-flux variables. In detail, the discretized Gauss constraints is given by
 \begin{equation}\label{disgauss}
 G_v:=\sum_{b(e)=v}X_e-\sum_{t(e')=v}h_{e'}^{-1}X_{e'}h_{e'}\approx0.
 \end{equation}
The discretized simplicity constraints include two sets. The first set is referred to as the edge-simplicity constraint $S^{IJKL}_e\approx0$, which takes the form \cite{Bodendorfer:Qu}\cite{Bodendorfer:SgI}
\begin{equation}
\label{simpconstr}
S_e^{IJKL}\equiv X^{[IJ}_e X^{KL]}_e\approx0, \ \forall e\in \Gamma.
\end{equation}
The second set is referred to as the vertex-simplicity constraint $S^{IJKL}_{v,e,e'}\approx0$, which can be expressed as \cite{Bodendorfer:Qu}\cite{Bodendorfer:SgI}
\begin{equation}\label{simpconstr2}
\quad S_{v,e,e'}^{IJKL}\equiv X^{[IJ}_e X^{KL]}_{e'}\approx0,\ \forall e,e'\in \Gamma, s(e)=s(e')=v.
\end{equation}
The symplectic structure of the discrete phase space can be given by the Poisson algebra between the elementary variables $(h_e, X^{IJ}_e)$, which reads
 \begin{eqnarray}
 &&\{h_e, h_{e'}\}=0,\quad\{h_e, X^{IJ}_{e'}\}=\delta_{e,e'}\frac{\kappa}{a^{D-1}} \frac{d}{dt}(e^{\lambda\tau^{IJ}}h_e)|_{\lambda=0}, \\\nonumber
 && \{X^{IJ}_e, X^{KL}_{e'}\}=\delta_{e,e'}\frac{\kappa}{a^{D-1}}(\delta^{IK}X_e^{JL}+\delta^{JL }X^{IK}_e-\delta^{IL}X_e^{JK}-\delta^{JK}X_e^{ IL}).
 \end{eqnarray}
 By using these Poisson algebras, one can verify that the Gaussian constraint induces the $SO(D+1)$ gauge transformation in $SO(D+1)$ Yang-Mills theory exactly, and the edge-simplicity constraint generates the transformation
 \begin{equation}\label{esimtrans}
 \{X_e^{[IJ}X_e^{KL]}, h_e\}= 2X_e^{[IJ}\{X_e^{KL]}, h_e\}=\frac{-2\kappa}{a^{D-1}} X_e^{[IJ}(\tau^{KL]}h_e).
\end{equation}
Moreover, one can calculate the Poisson algebra amongst the discretized Gauss constraints, edge-simplicity constraints and vertex-simplicity constraints. It has been shown that $G_v\approx0$ and $S_e\approx0$ form a first class constraint system, with the weakly commutative algebras
\begin{eqnarray}
\label{firstclassalgb}
\{S_e, S_e\}\propto S_e\,,\,\, \{S_e, S_v\}\propto S_e,\,\,\{G_v, G_v\}\propto G_v,\,\,\{G_v, S_e\}\propto S_e,\,\,\{G_v, S_v\}\propto S_v, \quad s(e)=v,
\end{eqnarray}
where the Poisson algebra within $G_v\approx0$ are isomorphic to the $so(D+1)$ algebra.
However, the algebras among the vertex-simplicity constraint are the problematic ones, with the open anomalous Poisson brackets \cite{Bodendorfer:2011onthe}
\begin{eqnarray}
\label{anomalousalgb}
\{S_{v,e,e'},S_{v,e,e''}\}\propto \emph{anomaly terms}
\end{eqnarray}
where the $ ``\emph{anomaly terms}''$ are not proportional to any of the existing constraints in the phase space.
Different with that in connection phase space, the anomalous Poisson algebra of the vertex simplicity constraint in holonomy-flux phase space leads that the constraint system is no longer first class. Thus, the gauge reduction in holonomy-flux phase space should not been a simple copy of the corresponding reduction in continuum phase space. The main obstacle to proceed the gauge reduction in holonomy-flux phase space is that how to deal with the anomaly of vertex simplicity constraint, so that the gauge degrees of freedom can be reduced correctly. This problem is solve by considering the generalized twisted geometric parametrization of the holonomy-flux phase space, with the twisted geometry covering the degrees of freedom of the Regge geometries so that it can get back to the connection phase space in some continuum limit \cite{PhysRevD.103.086016}. Let us give a brief review of this parametrization as follow.

From now on, our discussions in this article are completely focusing on a graph $\Gamma$ whose dual lattice gives a partition of $\sigma$ constituted by $D$-dimensional polytopes. The elementary edges in $\Gamma$ refers to such kind of edges which only pass through one $(D-1)$-dimensional face in the dual lattice of $\Gamma$. The holonomy-flux phase space associated to the give graph $\Gamma$ is given by $\times_{e\in \Gamma}T^\ast SO(D+1)_e$, with $e$ being the elementary edges of $\Gamma$. Then, the edge simplicity constraint surface in $\times_{e\in \Gamma}T^\ast SO(D+1)_e$ that we are interested in is given by \cite{PhysRevD.103.086016}
\begin{equation}
\times_{e\in \Gamma}T_{\text{s}}^\ast SO(D+1)_e:=\{(h_e,X_e)\in \times_{e\in \Gamma}T^\ast SO(D+1)_e|X_{e}^{[IJ}X_{e}^{KL]}=0\}.
\end{equation}
Without loss of generality, we can focus on the edge simplicity constraint surface $T_{\text{s}}^\ast SO(D+1)_e$ related to one single elementary edge $e\in\Gamma$.
This space can be parametrized by using the generalized twisted-geometry variables
 \begin{equation}
 (V_e,\tilde{V}_e,\xi_e, \eta_e,\bar{\xi}_e^\mu)\in P_e:=Q^e_{D-1}\times Q^e_{D-1}\times T^*S_e\times SO(D-1)_e,
 \end{equation}
 where $\eta_e\in\mathbb{R}$, $Q^e_{D-1}:=SO(D+1)/(SO(2)\times SO(D-1))$ is the space of unit bi-vectors $V_e$ or $\tilde{V}_e$ with $SO(2)\times SO(D-1)$ is the maximum subgroup fixing the bi-vector $\tau_o:=2\delta_1^{[I}\delta_2^{J]}$ in $SO(D+1)$, $\xi_e\in [-\pi,\pi)$, $e^{\bar{\xi}_e^\mu\bar{\tau}_\mu}:=\bar{u}_e$, and $\bar{\tau}_\mu$ with $\mu\in\{1,...,\frac{(D-1)(D-2)}{2}\}$ is the basis of the Lie algebra of the subgroup $SO(D-1)$ fixing both $\delta_1^{I},\delta_2^{J}$ in $SO(D+1)$. To capture the intrinsic curvature, we specify one pair of the $SO(D+1)$ valued Hopf sections $u_e:=u(V_e)$ and $\tilde{u}_e:=\tilde{u}( \tilde{V}_e)$ which satisfies $V_e=u_e\tau_ou_e^{-1}$ and $\tilde{V}_e=-\tilde{u}_e\tau_o\tilde{u}_e^{-1}$. Then, the parametrization associated with each edge is given by the map
\begin{eqnarray}\label{para}
(V_e,\tilde{V}_e,\xi_e,\eta_e,\bar{\xi}_e^\mu)\mapsto(h_e,X_e)\in T_{\text{s}}^\ast SO(D+1)_e:&& X_e=\frac{1}{2}\eta_e V_e=\frac{1}{2}\eta_eu(V_e)\tau_ou(V_e)^{-1}\\\nonumber
&&h_e=u(V_e)\,e^{\bar{\xi}_e^\mu\bar{\tau}_\mu}e^{\xi_e\tau_o}\,\tilde{u}(\tilde{V}_e)^{-1}.
\end{eqnarray}
 Now we can get back to the discrete phase space of all dimensional LQG on the whole graph $\Gamma$, which is just the Cartesian product of the discrete phase space on each single edge of $\Gamma$. Then, the twisted geometry parametrization of the discrete phase space on one copy of the edge can be generalized to that of the whole graph $\Gamma$ directly. Furthermore, the twisted geometry parameters $(V_e,\tilde{V}_e,\xi_e, \eta_e)$ take the interpretation of the discrete geometry describing the dual lattice of $\Gamma$, which can be explained explicitly as follows. We first note that $\frac{1}{2}\eta _e V_e$ and $\frac{1}{2}\eta _e \tilde{V}_e$ represent the area-weighted outward normal bi-vectors of the $(D-1)$-face dual to $e$ in the perspective of source and target points of $e$ respectively, with $\frac{1}{2}\eta _e$ being the dimensionless area of the $(D-1)$-face dual to $e$. Then, the holonomy $h_e=u_e(V_e)\,e^{\bar{\xi}_e^\mu\bar{\tau}_\mu}e^{\xi_e\tau_o}\,\tilde{u}^{-1}_e(\tilde{V}_e)$ takes the interpretation that it rotates the inward normal $-\frac{1}{2}\eta _e\tilde{V}_e$ of the (D-1)-face  dual to $e$ in the perspective of the the target point of $e$, into the outward normal $\frac{1}{2}\eta _e{V}_e$ of the (D-1)-face  dual to $e$ in the perspective of the source point of $e$, wherein $u_e(V_e)$ and $\tilde{u}_e(\tilde{V}_e)$ capture the contribution of intrinsic curvature, and $e^{\xi_e\tau_o}$ captures the contribution of extrinsic curvature to this rotation. Moreover, $\bar{u}_e=e^{\bar{\xi}_e^\mu\bar{\tau}_\mu}$ are some redundant degrees of freedom in the reconstruction of the discrete geometry, and it also contains the gauge degrees of freedom with respect to edge-simplicity constraint.
Then, beginning with the twisted geometry parameter space $P_\Gamma=\times_{e\in\Gamma}P_e, P_e:=Q^e_{D-1}\times Q^e_{D-1}\times T_e^*S\times SO(D-1)_e$ related to $\Gamma$, the gauge reduction with respect to the kinematic constraints---Gauss constraint and simplicity constraints---can be done by the guiding of their geometrical meaning in Regge geometry in the subset with $\eta_e\neq 0$. Up to a double-covering symmetry, we firstly reduce the $SO(D-1)_e$ fibers for each edge $e$ to get the phase space $\check{P}_\Gamma:=\times_{e\in \Gamma}\check{P}_e$ with $\check{P}_e:=Q^e_{D_1}\times Q^e_{D-1}\times T^*S_e^1$. Then, the discretized Gauss constraint \eqref{disgauss} can be imposed to give the reduced phase space
 \begin{equation}
\check{H}_\Gamma:=\check{P}_\Gamma/\!/SO(D+1)^{V(\Gamma)}=\left(\times_{e\in\Gamma} T^\ast S_e^1\right)\times \left(\times_{v\in\Gamma} \mathfrak{P}_{\vec{\eta}_v}\right)
\end{equation}
 with $V(\Gamma)$ being the number of the vertices in $\Gamma$ and
 \begin{equation}
 \mathfrak{P}_{\vec{\eta}_v}:=\{(V_{e_1}^{IJ},...,V_{e_{n_v}}^{IJ})\in \times_{e\in\{e_v\}}Q_{D-1}^{e}| G_{v}=0 \}/SO(D+1),
 \end{equation}
where we re-oriented the edges linked to $v$ to be out-going at $v$ without loss of generality, $\{e_v\}$ represents the set of edges beginning at $v$ with $n_v$ being the number of elements in $\{e_v\}$, and $G_{v}=\sum_{\{e_v\}}\eta_{e_v}V_{e_v}^{IJ}$ here. Further, we solve the vertex simplicity constraint equation \eqref{simpconstr} in the reduced phase space $\check{H}_\Gamma$ and get the final generalized twisted geometric space $\check{H}^{s}_\Gamma=\left(\times_{e\in\Gamma} T^\ast S_e^1\right)\times \left(\times_{v\in\Gamma} \mathfrak{P}^{s}_{\vec{\eta}_v}\right)$ with $\mathfrak{P}^{s}_{\vec{\eta}_v}:=\mathfrak{P}_{\vec{\eta}_v}|_{S_v=0}$. It has been shown that the generalized twisted geometry in the space $\check{H}^{s}_\Gamma$ is consistent with the Regge geometry on the spatial D-manifold $\sigma$ if the shape match condition in the D-polytopes' gluing process is considered, which means the gauge reduction scheme in the parametrization space captures the correct physical degrees of freedom of all dimensional LQG in kinematical level. Thus, based on this twisted geometry parametrization, one can conclude that, in order to get correct kinematical physical degrees of freedom, the anomalous vertex should be treated as a second class constraint while the Gauss constraint and edge simplicity constraint are treated as first class constraint in discrete and quantum theory of all dimensional LQG. The reduction procedures can be roughly illustrated as follows \cite{PhysRevD.103.086016}.
\begin{equation}\label{flowchart111}
\xymatrix{}
\xymatrix@C=1.5cm{
\times_{e\in\Gamma}T^\ast SO(D+1)_e
\ar[r]_{\ \  \  \ \text{(i)}}&
\times_{e\in\Gamma}\check{P}_e\ar[r]_{\text{(ii)}} & \check{H}_\Gamma^{}
\ar[r]_{\text{(iii)}}&
\check{H}_\Gamma^{s} },
\end{equation}
where  the symplectic reductions with respect to edge simplicity constraint and Gaussian constraint are proceeded in step (i) and (ii) respectively, and in step (iii) the vertex  simplicity constraint equation is solved.

\subsection{On the construction of the gauge invariant variables with respect to simplicity constraint}\label{gaugesim}
The symplectic reductions lead to the reduced phase space coordinatized by the gauge invariant variables, hence it is necessary to give the explicit expressions of the gauge invariant variables with respect to simplicity constraint. In fact, the gauge invariant variables with respect to simplicity constraint can be constructed by two schemes, which are referred to as the gauge averaging scheme and the gauge  fixing scheme respectively.

In the gauge averaging scheme, one need to consider the gauge averaging operation with respect to edge-simplicity constraint in the holonomy-flux phase space, which can be proceeded based on the twisted geometry parametrization. Let us focus on the constraint surface defined by edge-simplicity constraint in the phase space $T^\ast SO(D+1)_e$ associated to one single elementary edge $e$ of $\Gamma$. Based on the twisted geometry parametrization, we note that the gauge transformation induced by edge-simplicity constraint on the edge-simplicity constraint surface is given by
 \begin{eqnarray}\label{edgesimtrans2}
 \{X_e^{[IJ}X_e^{KL]}, h_e\}&=& 2X_e^{[IJ}\{X_e^{KL]}, h_e\}
 \propto \eta_eV_e^{[IJ}(\tau^{KL]}u_e\,e^{\bar{\xi}_e^\mu\bar{\tau}_\mu}e^{\xi_e\tau_o}\,\tilde{u}_e^{-1}) \\\nonumber&=&\eta_e(u_e\,(\bar{\tau}_e^{IJKL}e^{\bar{\xi}_e^\mu\bar{\tau}_\mu}) e^{\xi_e\tau_o}\,\tilde{u}_e^{-1})
\end{eqnarray}
and
\begin{equation}
\{X_e^{[IJ}X_e^{KL]}, X^{MN}_e\}=0,
\end{equation}
 where we defined $\bar{\tau}_e^{IJKL}:=V_e^ {[IJ}(u_e^{-1}\tau^{KL]}u_e)\in so(D-1)$. It easy to see that the edge simplicity constraint induce the transformation of the component $e^{\bar{\xi}_e^\mu\bar{\tau}_\mu}\in SO(D-1)$ in the parametrization of $h_e$, and the flux is gauge invariant with respect to edge-simplicity constraint on the constraint surface defined by edge-simplicity constraint. Thus, we only need to focus on the gauge reduction of holonomies. 
 To achieve this goal, let us consider the averaging operation $\mathbb{P}_{\text{S}}$ with respect to the gauge transformation generated by the edge-simplicity constraint in the holonomy-flux phase space, whose infinitely small transformation is given by \eqref{edgesimtrans2}. Then, the action of $\mathbb{P}_{\text{S}}$ on the holonomy and flux in the constraint surface defined by edge-simplicity constraint can be given as
\begin{equation}\label{qgr0}
\mathbb{P}_{\text{S}}\circ h_e :=\int_{SO(D-1)}d\bar{g}\left(u_ee^{\xi^o\tau_o}(\bar{g}e^{\bar{\xi}^\mu_e\bar{\tau}_\mu})\tilde{u}_e^{-1}\right) =h^{s}_e,
\end{equation}
\begin{equation}\label{qgr00}
\mathbb{P}_{\text{S}}\circ X_e =X_e,
\end{equation}
where we used that $h_e=u_ee^{\xi\tau_o}e^{\bar{\xi}^\mu_e\bar{\tau}_\mu}\tilde{u}_e^{-1}$, $\bar{g}\in SO(D-1)\subset SO(D+1)$, and $h^{s}_e$ is the simplicity reduced holonomy defined by
\begin{equation}
h^{s}_e=u_ee^{\xi\tau_o}\mathbb{I}^s\tilde{u}_e^{-1},
\end{equation}
where $(\mathbb{I}^s)^I_{\ J}:=(\delta_1)^I(\delta_1)_J+(\delta_2)^I(\delta_2)_J$. Now, the gauge invariant variables with respect to simplicity constraint can be constructed by the gauge averaging scheme on the simplicity constraint surface, which take the formulations of the functions of $(h^s_e, X_e)$.

Recall the simplicity reduced connection $A^S_{aIJ}:= A_{aIJ}-\gamma\bar{K}_{aIJ}$ constructed in connection phase space, we can establish the following correspondence $A^S_{aIJ}$ and the simplicity reduced holonomy $h^s_e$,
  $$\xymatrix{}
\xymatrix@C=3.5cm{
(A_{aIJ},\pi^{bKL})
\ar[d]_{\textrm{(1)}} \ar[r]_{\textrm{regularization}}&(h_e,X_e) \ar[d]^{\textrm{(2)}}\\
(A^S_{aIJ},\pi^{bKL})|_{S^{abIJKL}=0}\ar[r]^{\textrm{ correspondence}}& (h^s_e,X_e)|_{S_e=0, S_v=0}}$$
where in steps (1) and (2) the symplectic reduction with respect to simplicity constraint are proceeded by gauge fixing and gauge averaging schemes respectively.
Though the simplicity reduced holonomy $h^s_e$ and the simplicity reduced connection $A^S_{aIJ}$ has above correspondence relation, $h^s_e$ is not the holonomy defined by $A^S_{aIJ}$. This can be seen by considering the continuous limit  of $h^s_e$, which reads
\begin{equation}\label{hseconlim}
h^s_e=u_ee^{\xi\tau_o}\mathbb{I}^s\tilde{u}_e^{-1} ={u}_ee^{\xi^o_e\tau_o}\mathbb{I}^s{u}_e^{-1}h_e^{\Gamma}\simeq ({u}_e\mathbb{I}^s{u}_e^{-1}+\beta K^{\perp}_e)(\mathbb{I}+\Gamma_e),
\end{equation}
where the appearance of $\mathbb{I}^s$ leads  that $h^s_e$ is not the holonomy defined by $A^S_{aIJ}$.
In fact, this un-consistency between $h^s_e$ and $A^S_{aIJ}$ comes from the difference of gauge reduction schemes proceeded in the connection phase space and the holonomy-flux phase space. Next, let us consider the gauge fixing scheme to construct the gauge invariant variables with respect to the simplicity constraint in holonomy-flux phase space.

 In order to introduce the gauge fixing scheme, let us first notice that the gauge invariant variables with respect to simplicity constraint  can be given as some functions $O(A^S_{aIJ}, \pi^{bKL})$ defined on the simplicity constraint surface in the connection phase space, with the gauge fixing being taken by choosing the gauge component $\bar{K}_{aIJ}=0$. Then, in the holonomy-flux phase space, the gauge invariant variables with respect to simplicity constraint can be constructed by regularizing the corresponding gauge invariant variables $O(A^S_{aIJ}, \pi^{bKL})$ in the connection phase space, which leads to the functions $O'(h_e, X_e)$ defined on the simplicity constraint surface. More explicitly, notice that $A^S_{aIJ}=A_{aIJ}-\gamma \bar{K}_{aIJ}$ and $\bar{K}_{aIJ}$ can be rewritten as a function of $(A_{aIJ}, \pi^{bKL})$ by using Eq.\eqref{KFD}, thus we have $O(A^S_{aIJ}, \pi^{bKL})=O\left(A^S_{aIJ}(A_{cMN}, \pi^{dOP}), \pi^{bKL}\right)$ and it can be regularized by smearing $(A_{aIJ}, \pi^{bKL})$ accordingly. Indeed, the scalar constraint \eqref{scalar1} in connection phase space is constructed based on the gauge fixing scheme, and its regularization and quantization lead to the constraint operator in all dimensional LQG. As we will see, the resulting operator will fail to be the correct scalar constraint operator in the $SO(D+1)$ LQG in which the edge-simplicity constraint is imposed strongly, while it can be generalized as a reasonable scalar constraint operator in another all dimensional LQG theory in which the edge-simplicity constraint is solved weakly.

%In fact, notice that the disappearance of $e^{\bar{\xi}^\mu_e\bar{\tau}_\mu}$ in $h^s_e$ not only reduces the gauge degrees of freedom of $\check{\xi}^\mu_e$ corresponding to $\bar{K}_{aIJ}$, but also the degrees of freedom of $\bar{\zeta}^\mu_e$ corresponding to some components of $\Gamma_{aIJ}$. Thus, $h^s_\alpha$ can not capture the physical degrees of freedom as $h_\alpha$.  This result introduce new obstacles to the construction of the scalar constraint operator in   the $SO(D+1)$ LQG.

\subsection{The quantum theory of the $SO(D+1)$ LQG}
\subsubsection{The Hilbert space and kinematic constraints}
The Hilbert space $\mathcal{H}$ of all dimensional LQG is given by the completion of the space of cylindrical functions on the quantum configuration space, which can be decomposed into the sectors --- the Hilbert spaces associated to graphs. For a given graph $\Gamma$ with $|E(\Gamma)|$ edges, the related Hilbert space is given by $\mathcal{H}_\Gamma=L^2((SO(D+1))^{|E(\Gamma)|}, d\mu_{\text{Haar}}^{|E(\Gamma)|})$. This Hilbert space associates to the classical phase space $\times_{e\in\Gamma}T^\ast SO(D+1)_e$ aforementioned. A basis of this space is given by the spin-network functions constructed on $\Gamma$ which are labelled by (1) an $SO(D+1)$ representation $\Lambda$ assigned to each edge of $\Gamma$; and (2) an intertwiner $i_v$ assigned to each vertex $v$ of $\Gamma$. Then, each basis state $\Psi_{\Gamma,{\vec{\Lambda}}, \vec{i}}(\vec{h})$, as a wave function on $\times_{e\in\Gamma}SO(D+1)_e$, can be given by
\begin{eqnarray}
\Psi_{\Gamma,{\vec{\Lambda}}, \vec{i}}(\vec{h}(A))\equiv \bigotimes_{v\in\Gamma}{i_v}\,\, \rhd\,\, \bigotimes_{e\in\Gamma} \pi_{\Lambda_e}(h_{e}(A)),
\end{eqnarray}
where $\vec{h}(A):=(...,h_e(A),...), \vec{\Lambda}:=(...,\Lambda_e,...), e\in\Gamma$, $\vec{i}:=(...,i_v,...), v\in\Gamma$ , $\pi_{\Lambda_e}(h_{e})$ denotes the matrix of holonomy $h_e$ associated to edge $e$ in the representation labelled by $\Lambda_e$, and $\rhd$ denotes the contraction of  the representation matrixes of holonomies with the intertwiners. Hence, the wave function $\Psi_{\Gamma,{\vec{\Lambda}}, \vec{i}}(\vec{h}(A))$ is simply the product of the functions on $SO(D+1)$, which are given by specified components of the holonomy matrices selected by the intertwiners at the vertices.
The action of the elementary operators---holonomy operator and flux operator---on the spin-network functions can be given as
\begin{eqnarray}
 \hat{ h}_{e}(A)\circ \Psi_{\Gamma,{\vec{\Lambda}}, \vec{i}}(\vec{h}(A)) &=& { h}_e(A) \Psi_{\Gamma,{\vec{\Lambda}}, \vec{i}}(\vec{h}(A)) \\\nonumber
  \hat{F}_e^{IJ}\circ\Psi_{\Gamma,{\vec{\Lambda}}, \vec{i}}(\vec{h}(A)) &=&-\mathbf{i}\hbar\kappa\beta R_e^{IJ}\Psi_{\Gamma,{\vec{\Lambda}}, \vec{i}}(\vec{h}(A))
\end{eqnarray}
where the holonomy operator acts by multiplying,  $R_{e}^{IJ}:=\text{tr}((\tau^{IJ}h_e)^{\text{T}}\frac{\partial}{\partial h_e})$  is the right
invariant vector fields on $SO(D+1)$ associated to the edge $e$, and $\text{T}$ denoting the transposition of the matrix.

Now one can proceed the quantum gauge reduction procedures with respect to Gaussian and simplicity constraints to obtain the kinematic physical Hilbert space. To achieve this goal, one needs to solve the kinematic constraints, including Gaussian constraint, edge-simplicity constraint and vertex-simplicity constraint in $\mathcal{H}$. Following the results given in Sec.\ref{sec2.2}, the Gaussian constraint and edge-simplicity constraint are imposed strongly and the corresponding solution space is spanned by the edge-simple and gauge invariant spin-network states, which are constructed by assigning simple representations of $SO(D+1)$ to edges and gauge invariant intertwiners to vertices of the associated graphes. Besides, the anomalous vertex simplicity constraints are imposed weakly and the corresponding weak solutions are given by the spin-network states labelled by the simple coherent intertwiners at vertices \cite{long2019coherent}. Specifically, a typical spin-network state labelled by the gauge invariant simple coherent intertwiners at vertices is given by
\begin{equation}\label{scinter}
\Psi_{\Gamma,\vec{N},\vec{\mathcal{I}}_{\text{s.c.}}}(\vec{h}(A))=\text{tr}(\otimes_{e\in\Gamma} \pi_{N_e}(h_e(A))\otimes_{v\in\Gamma}\mathcal{I}_v^{\text{s.c.}})
\end{equation}
where $\pi_{N_e}(h_e(A))$ denotes the representation matrix of $h_e(A)$ with $N_e$ being an non-negative integer labeling a simple representation of $SO(D+1)$, and $\vec{\mathcal{I}}_{\text{s.c.}}$ is defined by $\vec{\mathcal{I}}_{\text{s.c.}}:=(...,\mathcal{I}_v^{\text{s.c.}},...)$ with $\mathcal{I}_v^{\text{s.c.}}$ being the so-called gauge invariant simple coherent intertwiner labeling the vertex $v\in\Gamma$ \cite{long2019coherent}. More explicitly, the gauge invariant simple coherent intertwiner is defined as
 \begin{equation}
 \mathcal{I}_v^{\text{s.c.}}:=\int_{SO(D+1)}dg\otimes_{e: b(e)=v}\langle N_e,V_e|g
 \end{equation}
 where all the edges linked to $v$ are re-oriented to be outgoing at $v$ without loss of generality, the labels $V_e$ satisfies the classical vertex-simplicity constraint as
 \begin{equation}
 V_e^{[IJ}V_{e'}^{KL]}=0,\quad \forall\  b(e)=b(e')=v,
 \end{equation}
 and $|N_e,V_e\rangle$ is the Perelomov type coherent state of $SO(D+1)$ in the simple representation space labelled by $N_e$ \cite{Long:2020euh}, which satisfies
 \begin{equation}
\langle N_e,V_e| \tau^{IJ}|N_e,V_e\rangle=\mathbf{i}N_eV_e^{IJ}.
 \end{equation}
 By taking specific superpositions of the spin-network states labelled by the simple coherent intertwiners, the coherent states labelled by the twisted geometry parameters can be established, and it has been verified that these coherent states have well-behaved peakedness and Ehrenfest Properties \cite{Calcinari_2020,Long:2021xjm,Long:2021lmd,Long:2022cex}

With the Gaussian and simplicity constraints being solved,  the spatial geometric operators can be constructed based on the elementary operators in the kinematic physical Hilbert space \cite{long2020operators,Long:2020agv,Zhang:2015bxa}. For example, the $(D-1)$-area operator reads
 \begin{equation}
 \widehat{\text{Ar}}(S_e)=\sqrt{2\hat{F}_e^{IJ}\hat{F}_{e,IJ}},
 \end{equation}
 which measures the area of the $(D-1)$-dimensional face $S_e$ traversed by $e$ in the dual lattice of $\Gamma$.
The $D$-volume operator reads
 \begin{equation}\label{Vdef}
 \hat{V}(v,\square)=\sqrt[2D-2]{\hat{Q}_v},
 \end{equation}
  which measures the volume of the $D$-dimensional cell $\square$ dual to $v$ in the dual lattice of $\Gamma$, where $\hat{Q}_v$ is a polynomial of the flux operator $\hat{F}_e^{IJ}$ \cite{Bodendorfer:Qu}.

\subsubsection{The issue in the construction of scalar constraint operator}\label{scalaroperator}
Indeed, the strong imposition of the Gaussian and edge-simplicity constraint in quantum theory gives the gauge reduction of the quantum states based on gauge averaging scheme. An important result of this imposition is that the holonomy operator $\hat{h}_e$ acting in the strong solution space $\mathcal{H}^s$ of edge-simplicity constraint is equivalent to the simplicity reduced holonomy operator $\hat{h}^s_e$ \cite{Long:2022thb}, which is defined by
 \begin{equation}\label{simholo}
\widehat{h^{s}_e}:=\widehat{\mathbb{P}}_{\text{S}}\hat{h}_e\widehat{\mathbb{P}}_{\text{S}},
\end{equation}
where the projection operator $\widehat{\mathbb{P}}_{\text{S}}$ projects an arbitrary quantum state in $\mathcal{H}_\Gamma$ into  $\mathcal{H}^{s}_{\Gamma}$. It has been shown that the classical correspondence of  $\hat{h}^s_e$ is the simplicity reduced holonomy $h^s_e$, which can not capture the degrees of freedom of the spatial intrinsic curvature. This result leads that the standard strategy introduced in Ref.\cite{Long:2022thb} is fail to construct the scalar constraint operator. Let us explain this point as follows.

The regularization and quantization of the scalar constraint \eqref{scalar1} in $SO(D+1)$ LQG follows the standard strategy as that in the $SU(2)$ LQG, except the appearance of the additional term $C_{\text{D}}$.
 %By projecting the covariant derivation of $\pi^{aIJ}$ properly, the term $\frac{\beta^2}{\sqrt{\det(q)}}(\bar{K}_{bIK}E^{aI})(\bar{K}_{aJ}^{\ \ \, K}E^{bJ})$  is re-formulated as a $DF^{-1}D$ term composed by the connection variables \cite{Bodendorfer:Ha}. In fact, the $DF^{-1}D$ term is a rather complicated function of $A_{aIJ}$ and $\pi^{bIJ}$ so that its regularization and quantization are full of ambiguities \cite{Bodendorfer:Qu}.  However, the key issue is not the treatment of the $DF^{-1}D$ term when we consider the quantization of the scalar constraint. As we will see, the operator $\hat{C}_{\text{E}}$ corresponding to Euclidean term  lose its original geometric interpretation in the space $\bigoplus_{\gamma}\mathcal{H}^{s}_{\gamma}$, since the simplicity reduced holonomy which will appear in $\hat{C}_{\text{E}}$ can not give the curvatures correctly. Let us explain this point explicitly as follows.
Following the regularization and quantization procedures introduced in \cite{Bodendorfer:Qu}, the Euclidean term $C_{\text{E}}$ and Lorentzian term $C_{\text{L}}$ can be quantized directly, which leads to
\begin{equation}\label{CECL}
\hat{C}_{\text{E}}[N]=\lim_{\epsilon\to 0}\sum_{\square\in \mathfrak{P}}\hat{C}^\square_{\text{E}}[N],\quad \hat{C}_{\text{L}}[N]=\lim_{\epsilon\to 0}\sum_{\square\in \mathfrak{P}}\hat{C}^\square_{\text{L}}[N]
\end{equation}
with
\begin{equation}\label{CE}
\hat{C}^\square_{\text{E}}[N]:=N(v_\square)\cdot {}^\epsilon\!\left(\widehat{\frac{\pi^{[a|IK|}{\pi^{b]\ J}_{\ K}}}{\sqrt{\det(q)}}}\right)_{v_\square}\cdot(\hat{h}_{\alpha_{s_a,s_b}})_{[IJ]}
\end{equation}
and
\begin{eqnarray}\label{CL}
\hat{C}^\square_{\text{L}}[N]&:=&\frac{2(1+\gamma^2)}{(8\kappa^2\hbar^2\gamma^3)^2}N(v_\square)\cdot {}^\epsilon\!\left(\widehat{\frac{\pi^{[a|IK|}}{\sqrt[4]{\det(q)}}}\right)_{v_\square} \cdot\widehat{ (h_{s_a})}_{I}^{\ M}\left[\widehat{(h_{s_a}^{-1})}_{MK}, [\hat{C}_{\text{E}}[1],\hat{V}(v_\square,\epsilon)]\right]
\\\nonumber
&&\cdot{}^\epsilon\!\left(\widehat{\frac{{\pi^{b]JL}}}{\sqrt[4]{\det(q)}}}\right)_{v_\square}\cdot\widehat{(h_{s_b})}_{J}^{\ N}\left[\widehat{(h_{s_b}^{-1})}_{NK}, [\hat{C}_{\text{E}}[1],\hat{V}(v_\square,\epsilon)]\right],
\end{eqnarray}
where $N(x)$ is the lapse function, $\square$ denotes an elementary cell of the hyper-cubic partition $\mathfrak{P}$ of $\sigma$, $\epsilon$ represents the scale of $\square$, $v_\square$ is a vertex of $\square$, $\hat{V}(v_\square,\epsilon)$ is the volume operator of the hyper-cube containing $v_\square$ and characterized by $\epsilon$,  $s_a$ represents the edges of $\square$ based at $v_\square$, $\alpha_{s_a,s_b}$ represents the oriented loop based at $v_\square$ and $s_a, s_b$. Besides, the operator ${}^\epsilon\!\left(\widehat{\frac{\pi^{[a|IK|}{\pi^{b]\ J}_{\ K}}}{\sqrt{\det(q)}}}\right)_{v_\square}$ and ${}^\epsilon\!\left(\widehat{\frac{\pi^{aIK}}{\sqrt[4]{\det(q)}}}\right)_{v_\square} $ are constructed by regularizing and quantizing the factors $\frac{\pi^{aIK}{\pi^{b\ J}_K}}{\sqrt{\det(q)}}$ and $\frac{\pi^{aIK}}{\sqrt[4]{\det(q)}}$ respectively, with the regularization being compatible with the partition $\mathfrak{P}$ at $v_\square$, see more details in Ref.\cite{Bodendorfer:Qu}. The regularization and quantization of the term $C_{\text{D}}$ are similar to that of $C_{\text{E}}$ \cite{Bodendorfer:Qu}. Recall the explicit expression \eqref{CD} of $C_{\text{D}}$ given by Eqs.\eqref{F-1},\eqref{barDaIJ} and \eqref{DaIJ}, we have the smeared expressions
\begin{eqnarray}
&&{}^\epsilon\!\left(q \cdot F^{-1}_{aIJ,bKL}\right)\\\nonumber
&:=&\frac{1}{4(D-1)}{}^\epsilon\!\left(\sqrt{q}\pi_{aAC} \right) {}^\epsilon\!\left(\sqrt{q}\pi_{bBD}\right)\left({}^\epsilon\!\left(\sqrt{q}^{-1}\pi^{cEC}\right) {}^\epsilon\!\left(\sqrt{q}\pi_{cE}^{\ \ D}\right)-\delta^{CD}\right)\\\nonumber
&&\cdot\left(\delta^{AB}\delta^{K[I}\delta^{J]L}-2\delta^{LA}\delta^{B[I}\delta^{J]K}\right)
\end{eqnarray}
and
\begin{eqnarray}
&&{}^\epsilon\!\left(\sqrt{q}^{-3/2}\bar{D}^{aIJ}\right)\\\nonumber
&:=&\left({}^\epsilon\!\left(\bar{\delta}^I_{[K}\right){}^\epsilon\!\left(\bar{\delta}^J_{L]}\right)\delta^a_b +\frac{2}{D-1}\left({}^\epsilon\!\left(\sqrt{q}^{-1}\pi^{aM[I}\right) {}^\epsilon\!\left(\bar{\delta}^{J]}_{[K}\right) {}^\epsilon\!\left(\sqrt{q}\pi_{bL]M}\right)-{}^\epsilon\!\left(n^{[I}n_{[K}\right)\delta^{J]}_{L]}\delta^a_b \right)\right)\\\nonumber
&&\cdot{}^\epsilon\!\left(\sqrt{q}^{-3/2} D^{bKL}\right),
\end{eqnarray}
where
\begin{equation}
{}^\epsilon\!\left(\bar{\delta}^{J}_{K}\right)= \left(\delta^J_K-{}^\epsilon\!\left(n^Jn_K\right)\right),\quad  {}^\epsilon\!\left(n^In_J\right)=\frac{1}{D-1}\left( {}^\epsilon\!\left(\sqrt{q}^{-1}\pi^{aKI}\right){}^\epsilon\!\left(\sqrt{q}\pi_{aKJ}\right) -\delta^I_J\right),
\end{equation}
and
\begin{equation}
{}^\epsilon\!\left(\sqrt{q}^{-3/2}D^{aIJ}\right):={}^\epsilon\!\left(\sqrt{q}^{-3/2}\pi^{b[I}_{\ \ \ K}\right){}^\epsilon\!\left(\mathcal{D}_b\pi^{a|K|J]}\right)
\end{equation}
with
\begin{equation}
{}^\epsilon\!\left(\mathcal{D}_a\pi^{bAB}\right):=\left({}^\epsilon\!\left(\pi^b(v_2) \right)-h_{s_a}^{-1}\cdot{}^\epsilon\!\left(\pi^b(v_1) \right)\cdot h_{s_a}\right)^{AB},\quad s(s_a)=v_1, \quad t(s_a)=v_2.
\end{equation}
%\begin{equation}
%\bar{D}^{aIJ}:=\left(\delta^a_b\bar{\delta}^I_{[K}\bar{\delta}^J_{L]}+\frac{2}{D-1}(\pi^{aM[I} \bar{\delta}^{J]}_{[K} \pi_{bL]M}-\delta^a_b\delta^{[I}_{[K}n^{J]}n_{L]})\right)D^{bKL}
%\end{equation}
The smeared factors ${}^\epsilon\!\left(\sqrt{q}\pi_{aIJ}\right) $, ${}^\epsilon\!\left(\sqrt{q}^{-3/2}\pi^{aKI}\right)$ and ${}^\epsilon\!\left(\sqrt{q}^{-1}\pi^{aKI}\right)$ can be quantized as the corresponding operators composed by holonomy and flux operators, see the details in Ref.\cite{Bodendorfer:Qu}.
Then, by quantizing each term in the smeared version of $C_{\text{D}}$, one can get the operator $\hat{C}_{\text{D}}[N]$ if we neglect the order of operators, which reads
\begin{equation}\label{CD}
\hat{C}_{\text{D}}[N]=\lim_{\epsilon\to 0}\sum_{\square\in \mathfrak{P}}\hat{C}^\square_{\text{D}}[N]
\end{equation}
with
\begin{equation}\label{CD1}
\hat{C}^\square_{\text{D}}[N]:=4N(v_\square){}^\epsilon\!\widehat{\left(\sqrt{q}^{-3/2}\bar{D}^{aIJ}\right)}_{v_\square}\cdot  {}^\epsilon\!\widehat{\left(q \cdot F^{-1}_{aIJ,bKL}\right)}_{v_\square}\cdot {}^\epsilon\!\widehat{\left(\sqrt{q}^{-3/2}\bar{D}^{bKL}\right)}_{v_\square}.
\end{equation}
Now, the scalar constraint operator $\hat{C}[N]$ in $SO(D+1)$ LQG is given by
\begin{equation}\label{scalarop}
\hat{C}[N]=\hat{C}_{\text{E}}[N]+\hat{C}_{\text{L}}[N]+\hat{C}_{\text{D}}[N]
\end{equation}
with $\hat{C}_{\text{E}}[N]$, $\hat{C}_{\text{L}}[N]$ and $\hat{C}_{\text{D}}[N]$ are given by Eqs. \eqref{CECL} and \eqref{CD}.

Notice that the operator ${}^\epsilon\!\left(\widehat{\frac{\pi^{[a|IK|}{\pi^{b]\ J}_{\ K}}}{\sqrt{\det(q)}}}\right)_{v_\square}$ in $\hat{C}_{\text{E}}[N]$ is a polynomial of $(\hat{V}(v_\square,\epsilon))^{1+x}$ and $\hat{h}_{s_a}(\hat{V}(v_\square,\epsilon))^{1+x}\hat{h}^{-1}_{s_a}$ with $x>-1$, thus it is commutative with $\widehat{\mathbb{P}}_{\text{S}}$. Then, consider a state $|\phi\rangle \in \bigoplus_{\Gamma}\mathcal{H}^{s}_{\Gamma}$ which satisfies
\begin{equation}
\widehat{\mathbb{P}}_{\text{S}}|\phi\rangle=|\phi\rangle,
\end{equation}
we have
\begin{equation}
\langle\phi|\hat{C}_{\text{E}}[N]|\phi'\rangle=\langle\phi|\widehat{\mathbb{P}}_{\text{S}} \hat{C}_{\text{E}}[N]\widehat{\mathbb{P}}_{\text{S}}|\phi' \rangle=\langle\phi|\hat{C}^s_{\text{E}}[N]|\phi'\rangle,
\end{equation}
where we defined
 \begin{equation}\label{CSE}
\hat{C}^s_{\text{E}}[N]:=\lim_{\epsilon\to 0}\sum_{\square\in \mathfrak{P}}\hat{C}^{s,\square}_{\text{E}}[N]
\end{equation}
with
\begin{equation}
\hat{C}^{s,\square}_{\text{E}}[N]:=N(v_\square)\cdot {}^\epsilon\!\left(\widehat{\frac{\pi^{[a|IK|}{\pi^{b]\ J}_{\ K}}}{\sqrt{\det(q)}}}\right)_{v_\square}\cdot(\widehat{h^s}_{\alpha_{s_a,s_b}})_{[IJ]},
\end{equation}
which is given by replacing the holonomy operator $\hat{h}_{\alpha_{s_a,s_b}}$ in $\hat{C}_{\text{E}}[N]$ by the simplicity reduced one $\widehat{h^s}_{\alpha_{s_a,s_b}}$. By this we can conclude that $\hat{C}_{\text{E}}[N]$ is equivalent to $\hat{C}^{s}_{\text{E}}[N]$ in the space $\bigoplus_{\Gamma}\mathcal{H}^{s}_{\Gamma}$. The key point of this result is that, if one consider the matrix element of $\hat{C}_{\text{E}}[N]$ in the space $\bigoplus_{\Gamma}\mathcal{H}^{s}_{\Gamma}$, the holonomy operator $\widehat{h}_{\alpha_{s_a,s_b}}$ reduces as the simplicity holonomy operator $\widehat{h^s}_{\alpha_{s_a,s_b}}$ and $\hat{C}_{\text{E}}[N]$ reduces as $\hat{C}^{s}_{\text{E}}[N]$. Note that  $\widehat{h^s}_{\alpha_{s_a,s_b}}$ corresponds to the  classical simplicity reduced holonomy $h^s_{\alpha_{s_a,s_b}}$, whose geometric interpretation is different with $h_{\alpha_{s_a,s_b}}$. Thus, we know that the action of $\hat{C}_{\text{E}}[N]$ in $\bigoplus_{\Gamma}\mathcal{H}^{s}_{\Gamma}$ can not reveal the physical meaning of the classical scalar constraint $C_{\text{E}}$ at quantum level.
Besides, the Eq.\eqref{CL} is also not the operator corresponding to $C_{\text{L}}$, since its definition relies on the operator $\hat{C}_{\text{E}}[1]$.

As we have explained in section \ref{gaugesim}, the scalar constraint operator \eqref{scalarop} is constructed as a gauge invariant variable with respect to simplicity constraint based on the gauge fixing scheme, which contradicts to the gauge averaging scheme used to proceed the gauge reduction of quantum states. Hence, it is reasonable that  the operator \eqref{scalarop} fail to be the correct scalar constraint operator in the Hilbert space $\mathcal{H}^s$ of $SO(D+1)$ LQG.
 In order to construct a correct scalar constraint operator in $SO(D+1)$ LQG, we proposed three new strategies in our previous work based on the simplicity reduced holonomy $h^s_e$ generated by gauge averaging scheme \cite{Long:2022thb}, which may provide feasible solutions to deal with the issues appearing in the construction of the scalar constraint operator. In the following part of this paper, we will turn to consider the weak coupling theory of all dimensional LQG based on the $U(1)^{\frac{D(D+1)}{2}}$ loop quantum theory. As we will see, the simplicity constraint in the weak coupling $U(1)^{\frac{D(D+1)}{2}}$ LQG will be treated by using the gauge fixing scheme, which would provide a new perspective of the construction of scalar constraint operator.

\section{The weak coupling $U(1)^{\frac{D(D+1)}{2}}$ LQG}

To establish  the $U(1)^{\frac{D(D+1)}{2}}$ LQG as our model of the of $SO(D+1)$ LQG in the weak-coupling limits, we first review the discrete phase spaces of $U(1)^{\frac{D(D+1)}{2}}$ and $SO(D+1)$, each coordinatized by their associated holonomy-flux variables. We will then show that there is indeed an asymptotically symplectic-morphism between the two phase spaces in the region where the holonomies approach identities. In this sense , the phase spaces of the two theories can indeed be asymptotically identified in the weak coupling limit region.

\subsection{ $SO(D+1)$ and $U(1)^{\frac{D(D+1)}{2}}$ flux-holonomy phase spaces }\label{bs}

Let us start from the $SO(D+1)$ case. Here we choose a basis $\{\tau_\alpha^{IJ}|\alpha\in\{1,2,...,\frac{D(D+1)}{2}\}\}$ for the Lie algebra $so(D+1)$, labeled by the index $\alpha$ in an ordering convenient for our later analysis:
\begin{equation}
\tau_\alpha^{IJ}=2\delta_1^{[I}\delta_{\alpha+1}^{J]},\ \ \text{for} \ \ \alpha\in\{1,...,D\},
\end{equation}
\begin{equation}
\tau_\alpha^{IJ}=2\delta_2^{[I}\delta_{\alpha-D+2}^{J]},\ \ \text{for} \ \ \alpha\in\{D+1,...,2D-1\},
\end{equation}
\begin{equation}
.\\ .\\.
\end{equation}
\begin{equation}
\tau_\alpha^{IJ}=2\delta_D^{[I}\delta_{D+1}^{J]},\ \ \text{for} \ \ \alpha=\frac{D(D+1)}{2}.
\end{equation}
As usual, we have
\begin{equation}
\delta_\alpha^\beta=-\frac{1}{2}\text{tr}(\tau_\alpha\tau^\beta)=\frac{1}{2}\tau_\alpha^{IJ}\tau^\beta_{IJ},\  \text{and}\ \  \delta^{[I}_K\delta^{J]}_L=\frac{1}{2}\tau_\alpha^{IJ}\tau^\alpha_{KL}.
\end{equation}
Using the new index, the non-vanishing Poisson bracket \eqref{Poisson1} for connection phase space reads
\begin{equation}
\{{A}_{a}^{\alpha}(x),\pi^{b}_{\beta}(y)\}=\kappa\beta\delta_a^b\delta^\alpha_\beta\delta^{(D)}(x-y)
\end{equation}
with $A_a^{IJ}={A}_{a}^{\alpha}\tau_\alpha^{IJ}$ and $\pi^b_{KL}=\pi^{b}_{\beta}\tau^\alpha_{KL}$. Correspondingly, the Poisson algebra in $SO(D+1)$ the holonomy-flux phase space, on a specified graph $\gamma$, is given by
\begin{eqnarray}\label{paalpha}
&&\{{{h}}_e[{A}],{{X}}^\alpha_{e'}\}= \delta_{e,e'}\frac{\kappa}{a^{D-1}}\tau^\alpha{{h}}_e[{A}],\quad \{{{h}}_e[{A}],\tilde{{X}}^\alpha_{e'}\}= - \delta_{e,e'}\frac{\kappa}{a^{D-1}}{{h}}_e[{A}]\tau^\alpha,\\\nonumber
&&\{{{X}}^\alpha_{e},{{X}}^\beta_{e'}\}= \delta_{e,e'}\frac{\kappa}{a^{D-1}}{f^{\alpha\beta}}_\lambda{{X}}^\lambda_{e'},\quad \{\tilde{{X}}^\alpha_{e},\tilde{{X}}^\beta_{e'}\}= \delta_{e,e'}\frac{\kappa}{a^{D-1}}{f^{\alpha\beta}}_\lambda\tilde{{X}}^\lambda_{e'},
\end{eqnarray}
with ${{X}}^{IJ}_{e}\equiv {{X}}^\alpha_{e}\tau_\alpha^{IJ}$ and ${f^{\alpha\beta}}_\lambda$ being the $so(D+1)$ structure constants given by ${f^{\alpha\beta}}_\lambda\equiv -\text{tr}(\tau^\alpha\tau^\beta\tau_\lambda)$.

Next, let us look into the $U(1)^{\frac{D(D+1)}{2}}$ case. The basis for the Lie algebra of the group $U(1)^{\frac{D(D+1)}{2}}$ is simply given by $\{\underline{\tau}^\alpha=\mathbf{i}|\alpha\in\{1,2,...,\frac{D(D+1)}{2}\}\}$ consisting of the $\frac{D(D+1)}{2}$ copies of $U(1)$ generator. The corresponding connection phase space will then be coordinatized by the canonical conjugate pairs $({A}_{a}^{\alpha}(x),\pi^{b}_{\beta}(y))$.
The holonomy-flux phase space associated to the same graph $\gamma$ chosen above can be now prescribed by the following. The holonomy over an oriented curve $e\in\Sigma$ is analogously defined by
\begin{equation}
\underline{h}_{e}^\alpha[{A}]\equiv e^{\mathbf{i}\int_{e}{A}_a^\alpha dx^a}.
\end{equation}
The $U(1)^{\frac{D(D+1)}{2}}$ flux variables for $\underline{\pi}^{a}_\beta$ can also be defined over an oriented $(D-1)$-surface. Just as in the $SO(D+1)$ case,  for the phase space,  the $(D-1)$-surface $S_e$ is dual to an edge $e$ of $\gamma$, and the flux variable over $S_e$ is given by
\begin{equation}
\underline{F}_\beta(e)\equiv\frac{1}{\beta } \int_{S_e}\epsilon_{a_1a_2...a_D}{{\pi}}^{a_1}_\beta d\sigma^{a_2}\wedge...\wedge d\sigma^{a_D}.
\end{equation}
The symplectic structure of the $U(1)^{\frac{D(D+1)}{2}}$  holonomy-flux phase space, coordinated by $\{ \underline{h}_{e}^\alpha,{F}_\beta(e)\}$, is determined by the only non-vanishing Poisson brackets of
\begin{equation}
\label{originalU(1)3}
\{\underline{h}_{e}^\alpha,\underline{F}^\beta(e)\}=\delta^{\alpha\beta}\mathbf{i}\kappa \underline{h}_{e}^\alpha,
\end{equation}
where $\epsilon(e',S_e)$ is the sign of the relative orientation between the given $e'$ and $S_e$ if they are dual to each other, and is zero otherwise, $\Gamma(S_e)$ has been adapted to $S_e$ by adding pseudo vertices such that they only intersect at the vertices of the former.

\subsection{Re-parametrization}\label{rp}

 In this subsection, we show a privileged parametrization of the $SO(D+1)$ holonomy-flux phase space using the coordinates of the $U(1)^{\frac{D(D+1)}{2}}$ holonomy-flux phase space. The parametrization as a map between the two phase spaces preserves the Poisson structure in the region of the weak coupling limit. This serves as the foundation of using the $U(1)^{\frac{D(D+1)}{2}}$ holonomy-flux formulation as an approximation for the loop representation of $SO(D+1)$ connection formulation of GR under the limit.

Based on the same graph $\Gamma$, it is clear that the $SO(D+1)$ and $U(1)^{\frac{D(D+1)}{2}}$ holonomy-flux phase spaces  have the same dimensionality. Hence it is possible to parametrize the $SO(D+1)$ holonomy-flux using the $U(1)^{\frac{D(D+1)}{2}}$ holonomy-flux variables.
Referring to the expressions \eqref{X} and \eqref{tildeX}, we specifically set the re-parametrization to be the map given by
\begin{eqnarray}\label{repa}
\underline{h}_{e}^\alpha[{A}]&\mapsto& h_{e}[A]:\quad \quad \quad h_{e}[A]\equiv\underline{\tilde{h}}_{e}[{A}]:=\exp\left(\frac{\sum_{\alpha}\left(\underline{h}_{e}^\alpha[{A}] -(\underline{h}_{e}^i[{A}])^{-1}\right)\tau_\alpha}{2\mathbf{i}}\right),\\\nonumber
\underline{X}^\alpha_e&\mapsto&{X}^\alpha_e:\quad \quad \quad   \quad \quad   \quad {X}^\alpha_e\tau_\alpha\equiv\underline{X}^\alpha_e\tau_\alpha:=\underline{\tilde{h}}^{1/2}_{e}[A]\tau_\alpha \underline{\tilde{h}}^{-1/2}_{e}[A]{\underline{Y}}^\alpha_e
\end{eqnarray}
where
\begin{eqnarray}
\underline{\tilde{h}}^{1/2}_{e}[A]:=\exp\left(\frac{\sum_{\alpha}\left(\underline{h}_{e}^\alpha[{A}] -(\underline{h}_{e}^\alpha[{A}])^{-1}\right)\tau_i}{4\mathbf{i}}\right)
\end{eqnarray}
and the function $\exp(\,\,): so(D+1)\to SO(D+1)$ should be understood as the exponential map of $so(D+1)$.
For the description of the flux variable in the frame of the target point of the edge, we again introduce
\begin{eqnarray}
 \tilde{\underline{X}}^\alpha_e\tau_\alpha:=-\underline{\tilde{h}}^{-1/2}_{e}[A]\tau_\alpha \underline{\tilde{h}}^{1/2}_{e}[A]{\underline{Y}}^\alpha_e.
\end{eqnarray}
Observe that the above implies the relation
\begin{equation}\label{relaXX}
\tilde{\underline{X}}^\alpha_e\tau_\alpha=-\underline{\tilde{h}}^{-1}_{e}[{A}]{\underline{X}}^\alpha_e \tau_\alpha\underline{\tilde{h}}_{e}[{A}] ,
\end{equation}
which is analogous to the relation \eqref{XXrelation}, and thus we may the interpret of the $U(1)^{\frac{D(D+1)}{2}}$ phase space functions $\underline{X}^\alpha_e$ and $\tilde{\underline{X}}^\alpha_e$ as the two descriptions of the same flux variables $X^\alpha_e$ and $\underline{X}^\alpha_e$ based on the two local frames for the fundamental $SO(D+1)$ theory associated to the source and target of the edge.

We now the crucial task of checking the Poisson-algebra consistency under the parametrization map \eqref{repa}, and see if the $SO(D+1)$ holonomy-flux algebra can truly be preserved in the $U(1)^{\frac{D(D+1)}{2}}$ phase space in the desired limits. Using the Poisson structures in the$U(1)^{\frac{D(D+1)}{2}}$ holonomy-flux phase space, it is straight forward to show that we have
\begin{equation}\label{hp0}
\{\underline{\tilde{h}}_{e}[{A}],\underline{\tilde{h}}_{e'}[{A}]\}=0,
\end{equation}
\begin{equation}\label{hp10}
\{\underline{\tilde{h}}_e[{A}],{\underline{X}}^\alpha_{e'}\}= - \delta_{e,e'}\frac{\mathbf{i}\kappa}{2a^{D-1}}\sum_{\beta}\text{tr}(\tau^\alpha\underline{\tilde{h}}^{1/2}_{e}[A] \tau_\beta\underline{\tilde{h}}^{-1/2}_{e}[A]) \underline{h}_e^\beta[{A}]\frac{\delta \underline{\tilde{h}}_{e}[{A}]}{\delta \underline{h}_e^\beta[{A}]},
\end{equation}
\begin{equation}\label{hp2}
\{\underline{\tilde{h}}_e[{A}],\tilde{\underline{X}}^\alpha_{e'}\}= \delta_{e,e'}\frac{\mathbf{i}\kappa}{2a^{D-1}}\sum_{\beta}\text{tr}(\tau^\alpha\underline{\tilde{h}}^{-1/2}_{e} [{A}] \tau_\beta\underline{\tilde{h}}^{1/2}_{e} [{A}]) \underline{h}_e^\beta[{A}]\frac{\delta \underline{\tilde{h}}_{e}[{A}]}{\delta \underline{h}_e^\beta[\mathcal{A}]},
\end{equation}
and
\begin{eqnarray}\label{hp3}
&&\{\tilde{\underline{X}}^\alpha_{e},\tilde{\underline{X}}^\beta_{e'}\}\\\nonumber
&=& \frac{1}{2}\delta_{e,e'}\text{tr}\left(\tau^\alpha\{\underline{\tilde{h}}^{-1/2}_{e} [A], \tilde{\underline{X}}^\beta_{e}\}\tau_\rho\underline{\tilde{h}}^{1/2}_{e} [{A}] \right)\underline{Y}^\rho_{e}+\frac{1}{2}\delta_{e,e'}\text{tr}\left(\tau^\alpha\underline{\tilde{h}}^{-1/2}_{e} [A]\tau_\rho \{\underline{\tilde{h}}^{1/2}_{e} [{A}] ,\tilde{\underline{X}}^\beta_{e}\}\right)\underline{Y}^\rho_{e}\\\nonumber
&&-\frac{1}{2}\delta_{e,e'}\text{tr}\left(\tau^\alpha\underline{\tilde{h}}^{-1/2}_{e} [A]\tau_\rho \underline{\tilde{h}}^{1/2}_{e} [{A}] \right)\left(\frac{1}{2}\text{tr}\left(\tau^\beta\{\underline{\tilde{h}}^{-1/2}_{e} [A], {\underline{Y}}^\rho_{e}\}\tau_\lambda\underline{\tilde{h}}^{1/2}_{e} [{A}] \right)\underline{Y}^\lambda_{e}\right)\\\nonumber
&&- \frac{1}{2}\delta_{e,e'}\text{tr}\left(\tau^\alpha\underline{\tilde{h}}^{-1/2}_{e} [A]\tau_\rho\underline{\tilde{h}}^{1/2}_{e} [{A}] \right)\left(\frac{1}{2}\text{tr}\left(\tau^\beta\underline{\tilde{h}}^{-1/2}_{e} [A]\tau_\lambda \{\underline{\tilde{h}}^{1/2}_{e} [{A}] ,{\underline{Y}}^\rho_{e}\}\right)\underline{Y}^\lambda_{e}\right),
\end{eqnarray}
wherein
\begin{equation}\label{hp1}
\{\underline{\tilde{h}}_e[{A}],{\underline{Y}}^\alpha_{e'}\}=\delta_{e,e'}\frac{\mathbf{i}\kappa}{a^{D-1}} \underline{h}_e^\alpha[{A}]\frac{\delta \underline{\tilde{h}}_{e}[{A}]}{\delta \underline{h}_e^\alpha[{A}]},\quad (\text{No summation over}\ \alpha).
\end{equation}
There should be no surprise that the Poisson algebra of the $SO(D+1)$ holonomy and fluxes defined by \eqref{repa} in the $U(1)^{\frac{D(D+1)}{2}}$ phase space does not coincide with the true  $SO(D+1)$ phase space algebra under the map. However, the coincidence occurs in the weak coupling limits. For a controlled  asymptotic analysis we introduce the parameter $\epsilon$ for the weak coupling limit, and study the correspondence under the matching conditions
\begin{eqnarray}
\label{match cond}
\underline{X}_e=X_e\,,\,\,\underline{h}^\alpha(e)=e^{\mathbf{i}\,\epsilon\, \theta^\alpha_e} \,\,\text{and}\,\,h(e)=e^{\,\epsilon\, \theta^\alpha_e\tau_\alpha}.
\end{eqnarray}
Under such conditions we immediately have
$$h_{e}[A]=\underline{\tilde{h}}_{e}[{A}]+\mathcal{O}(\epsilon^2).$$ This implies that, when \eqref{match cond} is satisfied , any $SO(D+1)$ phase space function $G(X_e ,h_{e})$ and the corresponding $U(1)^{\frac{D(D+1)}{2}}$ phase space function $G(\underline{X}_e, \underline{\tilde{h}}_{e})$ also agree to the same order
$$G(X_e ,h_{e})=G(\underline{X}_e, \underline{\tilde{h}}_{e})(1+\mathcal{O}(\epsilon^2)).$$
The  most important examples of these functions are the respective constraints governing the two theories. In either theory, the constraints act on the associated phase space via their Poisson brackets with the phase space coordinates. Due to the single differentiation operation involved, one expect the Poisson bracket $\{G, G'\}$ between any two functions $G$ and $G'$ to coincides between the two theories to the zeroth order of $\epsilon$.  This can be verified by looking into the $\epsilon^0$ contribution to the elementary $SO(D+1)$ loop algebra under the correspondence map.  For the true algebra in the $SO(D+1)$ phase space we have
\begin{eqnarray}\label{hp220}
&&\{{{h}}_e[{A}],{{X}}^\alpha_{e'}\}= \delta_{e,e'}\frac{\kappa}{a^{D-1}}\tau^\alpha+\mathcal{O}(\epsilon),\quad \{{{h}}_e[{A}],\tilde{{X}}^\alpha_{e'}\}= - \delta_{e,e'}\frac{\kappa}{a^{D-1}}\tau^\alpha+\mathcal{O}(\epsilon),\\\nonumber
&&\{{{X}}^\alpha_{e},{{X}}^\beta_{e'}\}= \delta_{e,e'}\frac{\kappa}{a^{D-1}}{f^{\alpha\beta}}_\lambda{{X}}^\lambda_{e'}+\mathcal{O}(\epsilon),\quad \{\tilde{{X}}^\alpha_{e},\tilde{{X}}^\beta_{e'}\}= \delta_{e,e'}\frac{\kappa}{a^{D-1}}{f^{\alpha\beta}}_\lambda\tilde{{X}}^\lambda_{e'}+\mathcal{O}(\epsilon).
\end{eqnarray}
Under the correspondence map, in $U(1)^{\frac{D(D+1)}{2}}$ phase space we have
\begin{eqnarray}\label{hp33}
&&\{\underline{\tilde{h}}_e[{A}],{\underline{X}}^\alpha_{e'}\}= \delta_{e,e'}\frac{\kappa}{a^{D-1}}\tau^\alpha+\mathcal{O}(\epsilon),\quad \{\underline{\tilde{h}}_e[{A}],\tilde{\underline{X}}^\alpha_{e'}\}= - \delta_{e,e'}\frac{\kappa}{a^{D-1}}\tau^\alpha+\mathcal{O}(\epsilon),\\\nonumber
&&\{{\underline{X}}^\alpha_{e},{\underline{X}}^\beta_{e'}\}= \delta_{e,e'}\frac{\kappa}{a^{D-1}}{f^{\alpha\beta}}_\lambda{\underline{X}}^\lambda_{e'}+\mathcal{O}(\epsilon), \quad\{\tilde{\underline{X}}^\alpha_{e},\tilde{\underline{X}}^\beta_{e'}\}= \delta_{e,e'}\frac{\kappa}{a^{D-1}}{f^{\alpha\beta}}_\lambda\tilde{\underline{X}}^\lambda_{e'}+\mathcal{O}(\epsilon).
\end{eqnarray}
Thus indeed, the brackets agree to the zeroth order as expected. It is in this sense, under the matching condition,  the Poisson algebra of the $SO(D+1)$ holonomy and fluxes defined via \eqref{repa} in the $U(1)^{\frac{D(D+1)}{2}}$ phase space coincides with the original one in the $SO(D+1)$ phase space, in the weak coupling limit $\epsilon\to 0$.  Finally, the parametrization \eqref{repa} is commutative with the re-orientation of the edges, so that we have
\begin{equation}
\underline{\tilde{h}}^{-1}_{e}[{A}]=\underline{\tilde{h}}_{e^{-1}}[{A}], \quad\tilde{\underline{X}}^\alpha_{e^{-1}}={\underline{X}}^\alpha_{e},\quad {\underline{X}}^\alpha_{e^{-1}}=\tilde{\underline{X}}^\alpha_{e}.
\end{equation}
Since the variables $\underline{\tilde{h}}_{e}[{A}]$, ${\underline{X}}^\alpha_{e}$ and $\tilde{\underline{X}}^\alpha_e$ in $U(1)^{\frac{D(D+1)}{2}}$ theory inherit the explicit structure of the corresponding variables of $SO(D+1)$ LQG, we may identify the $U(1)^{\frac{D(D+1)}{2}}$ phase space with the $SO(D+1)$ phase space in the weak coupling limit $\epsilon\to 0$, through the map given in \eqref{repa}.

\subsection{ Quantization of the $U(1)^{\frac{D(D+1)}{2}}$ holonomy-flux phase space: spin-network }

The above observations motivate us to apply the standard loop quantization method to the $U(1)^{\frac{D(D+1)}{2}}$ phase space to explore the weak coupling limits of $SO(D+1)$ loop quantum gravity. As we will see, the result is a $U(1)^{\frac{D(D+1)}{2}}$ loop quantum theory having a much simpler form than the full theory of $SO(D+1)$ loop quantum gravity.

The kinematic Hilbert space $\mathcal{K}$ of the $U(1)^{\frac{D(D+1)}{2}}$ loop quantum theory follows from the standard loop quantum representation of the holonomy-flux algebra with the gauge group of $U(1)^{\frac{D(D+1)}{2}}$. One way to identify a basis of the kinematic Hilbert space is to define the so-called charged holonomy $\underline{h}_{e,\vec{q}}[{A}]$ with a multiplet of integer charges $\{q^\alpha\}\equiv\vec{q}$ as
\begin{equation}
\underline{h}_{e,\vec{q}}[{A}]\equiv e^{\mathbf{i} q_\alpha\int_{e}{A}_a^\alpha dx^a}.
\end{equation}
Given a closed, oriented graph $\Gamma$ consisting of a set of edges $\{e_i\}$ meeting only at their end points, called the vertices, one may assign $\{\vec{q}_i\}$ to the edge $e_i\in\Gamma$ and thereby
define the graph holonomy $\underline{h}_{\Gamma,\{\vec{q}_i\}}$ as
\begin{equation}\label{grho}
\underline{h}_{\Gamma,\{\vec{q}_i\}}[{A}]\equiv \prod_{i}\underline{h}_{e_i,\vec{q}_i}[{A}].
\end{equation}
Note that, as in $SO(D+1)$ LQG, the kinematical Hilbert space $\mathcal{K}$ can be regarded as a union of the graph-dependent Hilbert  spaces $\mathcal{K}_{\Gamma}\equiv L^2\left((U(1)^{\frac{D(D+1)}{2}})^{|E(\Gamma)|}, d\mu_{\text{Haar}}^{|E(\Gamma)|}\right)$ on all possible graphs $\Gamma$ with each $U(1)^{\frac{D(D+1)}{2}}$ associated to an edge being thought as its holonomies. Here
%the Hilbert space $\mathcal{K}_{\gamma}$ over a graph $\gamma$ for the weak coupling theory consists of
$L^2\left((U(1)^{\frac{D(D+1)}{2}})^{|E(\Gamma)|}\right)$ is the space of square-integrable functions on $(U(1)^{\frac{D(D+1)}{2}})^{|E(\Gamma)|}$, and $d\mu_{\text{Haar}}^{|E(\Gamma)|}$ denotes the product of the Haar measure on $U(1)^{\frac{D(D+1)}{2}}$.
The $U(1)^{\frac{D(D+1)}{2}}$ kinematic Hilbert space $\mathcal{K}_{\Gamma}\equiv\text{Span}\{|c\rangle\}$ can be spanned by the basis of all the distinct charge network states and equipped with the inner product
\begin{equation}\label{inner}
\langle c| c' \rangle=\delta_{c,c'}
\end{equation}
with $c\equiv c(\Gamma,\{\vec{q}_i\})$.
 Note that the labeling $(\Gamma,\{\vec{q}_i\})$ to the charge network states is not unique, since one can always artificially change $\Gamma$ into $\Gamma'$ by adding trivial vertices and edges. To avoid this redundancy we will always label a charge network state by the corresponding oriented graph with the minimal number of edges.
In the Hilbert space $\mathcal{K}_{\Gamma}$, a holonomy operator acts as a multiplicative operator.
 A flux operator then acts as a differential operator such that
\begin{equation}
\underline{\hat{F}}^\beta(e)\cdot\underline{h}_{\Gamma,\{\vec{q}_i\}}[{A}]=\sum_{e'\in \Gamma(S_e)}\frac{1}{2}\hbar\kappa\epsilon(e',S_e)q^\beta_{e'}\underline{h}_{\Gamma,\{\vec{q}_i\}}[{A}].
\end{equation}

The Hilbert space $\mathcal{K}$ of this $U(1)^{\frac{D(D+1)}{2}}$ theory also has a coherent state basis. For the given graph $\Gamma$, one has ${\underline{H}}:=\{\underline{H}_e=\{\underline{H}_e^\alpha\}|e\in\Gamma\}$ which coordinatizes the holonomy-flux phase space $(T^\ast U(1)^{\frac{D(D+1)}{2}})^{|E(\Gamma)|}$ with $\underline{H}_e^\alpha=e^{\mathbf{i}(\underline{\phi}^\alpha_e+\mathbf{i}\underline{Y}^\alpha_e )}$. The holonomy and flux in $(T^\ast U(1)^{\frac{D(D+1)}{2}})^{|E(\Gamma)|}$ can be given by $\underline{H}$ as
\begin{equation}
\underline{h}^\alpha_e(\underline{H})=e^{\mathbf{i}\underline{\phi}^\alpha_e}, \quad \underline{F}^\alpha_e(\underline{H})=a^{D-1}\underline{Y}^\alpha_e,
\end{equation}
 where  $\underline{Y}^\alpha_e$ can be regarded as the dimensionless flux in the $U(1)^{\frac{D(D+1)}{2}}$ theory, and $a$ is an arbitrary but fixed constant with the dimension of length. Then, the heat-kernel coherent states in this theory are given by
 \begin{equation}
\underline{\Psi}^{{t}}_{\Gamma,{\underline{H}}}({\underline{h}})=\prod_{e\in\Gamma} \underline{\Psi}^{t}_{\underline{H}_e}(\underline{h}_e)
\end{equation}
where ${\underline{h}}:=\{\underline{h}_e|e\in\Gamma\}$, and $\underline{\Psi}^{t}_{\underline{H}_e}(\underline{h}_e)$ denotes the heat-kernel coherent states for $U(1)^{\frac{D(D+1)}{2}}$ defined by
\begin{equation}
\underline{\Psi}^{t}_{\underline{H}_e}(\underline{h}_e):=\prod_{\alpha\in\{1,...,\frac{D(D+1)}{2}\}} \sum_{n_\alpha=-\infty}^{\infty}e^{-\frac{t}{2}n_\alpha^2}e^{\mathbf{i}n_\alpha(\underline{\phi}^\alpha_e -\underline{\theta}^\alpha_e )}e^{-n_\alpha \underline{Y}^\alpha_e}
\end{equation}
such that $\underline{h}_e=\{\underline{h}_e^\alpha\}$ with $\underline{h}_e^\alpha=e^{\mathbf{i}\underline{\theta}^{\alpha}_e}$ and $t=\frac{\kappa\hbar}{a^{D-1}}$.

\subsection{Re-constructed operators and quantum algebras}

To extend the re-parametrization \eqref{repa} to quantum theory, one need to define the operators corresponding to the re-constructed $SO(D+1)$ holonomy-fluxes variables in $U(1)^{\frac{D(D+1)}{2}}$ loop quantum theory.
By construction, the re-constructed variables $\underline{\tilde{h}}_{e}[{A}]$, ${\underline{X}}^\alpha_{e}$ and $\tilde{\underline{X}}^\alpha_e$ in $U(1)^{\frac{D(D+1)}{2}}$ loop quantum theory can be directly quantized as $\hat{\underline{\tilde{h}}}_{e}[{A}]$, $\hat{{\underline{X}}}^\beta_e$ and $\hat{\tilde{\underline{X}}}^\beta_e$ respectively, which are defined by
\begin{eqnarray}\label{reop}
\hat{\underline{\tilde{h}}}_{e}[{A}]&:=& \exp\left(\frac{1}{2\mathbf{i}}\sum_{\alpha}\left(\underline{\hat{h}}_{e}^\alpha[{A}] -(\underline{\hat{h}}_{e} ^\alpha[{A}])^{-1}\right)\tau_\alpha\right),\\\nonumber
\hat{{\underline{X}}}^\beta_e&:=&-\frac{1}{4} \text{tr}(\tau^\beta \hat{\underline{\tilde{h}}}_{e}[{A}/2]\tau_\alpha\hat{\underline{\tilde{h}}}^{-1}_{e}[{A}/2]) \underline{\hat{Y}}^\alpha_e-\frac{1}{4} \underline{\hat{Y}}^\alpha_e\text{tr}(\tau^\beta \hat{\underline{\tilde{h}}}_{e}[{A}/2]\tau_\alpha\hat{\underline{\tilde{h}}}^{-1}_{e}[{A}/2])\\\nonumber
\hat{\tilde{\underline{X}}}^\beta_e&:=& \frac{1}{4} \text{tr}(\tau^\beta \hat{\underline{\tilde{h}}}_{e}^{-1}[{A}/2]\tau_\alpha\hat{\underline{\tilde{h}}}_{e}[{A}/2]) \underline{\hat{Y}}^\alpha_e+\frac{1}{4} \underline{\hat{Y}}^\alpha_e\text{tr}(\tau^\beta \hat{\underline{\tilde{h}}}_{e}^{-1}[{A}/2]\tau_\alpha\hat{\underline{\tilde{h}}}_{e}[{A}/2]).
\end{eqnarray}
The operators $\hat{{\underline{X}}}^\beta_e$ and $\hat{\tilde{\underline{X}}}^\beta_e$ are symmetric and hence admit self-adjoint extensions. Now, in order to verify that the $U(1)^{\frac{D(D+1)}{2}}$ loop quantum theory reveals the key quantum characters of $SO(D+1)$ LQG in the weak coupling limit, it is sufficient to show that the quantum algebras amongst the re-constructed $SO(D+1)$ holonomy-flux operators in the $U(1)^{\frac{D(D+1)}{2}}$ loop quantum theory are isomorphic to the corresponding Poisson algebras in $SO(D+1)$ holonomy-flux phase space in the weak coupling limit, up to the quantum parameter $\mathbf{i}\hbar$.  Notice that the weak coupling limit is given by small $\underline{\phi}^\alpha_e=\phi_e^\alpha$. Thus, to ensure our discussion only involves the weak coupling properties of the $U(1)^{\frac{D(D+1)}{2}}$ loop quantum theory, let us consider the normalized heat-kernel coherent states $\underline{\Phi}^{{t}}_{\underline{H}_e}(\underline{h}_e)$  in $U(1)^{\frac{D(D+1)}{2}}$ loop quantum theory, which are sharply peaked at the $U(1)^{\frac{D(D+1)}{2}}$ holonomy-flux phase space points with $\underline{\phi}^\alpha_e=\phi_e^\alpha$ being small. The coherent states $\underline{\Phi}^{{t}}_{\underline{H}_e}(\underline{h}_e)$ are composed by the heat-kernel coherent states of $U(1)$, and it has been proven that the coherent states $\underline{\Phi}^{{t}}_{\underline{H}_e}(\underline{h}_e)$ have well-behaved Ehrenfest properties. By this we mean that  the expectation values of the polynomials of the elementary
operators in the $U(1)^{\frac{D(D+1)}{2}}$ loop quantum theory, as well as the operators which are not polynomial functions of the elementary operators,
reproduce, to zeroth order in $t$, the values of the corresponding classical functions at the twisted
geometry space point where the coherent state is peaked. Then, it is straightforward to give
\begin{equation}
\frac{1}{\mathbf{i}\hbar}\langle \underline{\Phi}^{{t}}_{\underline{H}_e}| [\hat{\underline{\tilde{h}}}_{e}, \hat{\underline{\tilde{h}}}_{e}] |\underline{\Phi}^{{t}}_{\underline{H}_e}\rangle =\{\underline{\tilde{h}}_{e}, \underline{\tilde{h}}_{e}\}=\{h_{e}, h_{e}\}=0,
\end{equation}
\begin{equation}
\frac{1}{\mathbf{i}\hbar}\langle \underline{\Phi}^{{t}}_{\underline{H}_e}| [\hat{\underline{\tilde{h}}}_{e}, \hat{{\underline{X}}}^\beta_e] |\underline{\Phi}^{{t}}_{\underline{H}_e}\rangle=\{{\underline{\tilde{h}}}_{e}, {\underline{X}}^\beta_e\}+\mathcal{O}(t)
\end{equation}
and
\begin{equation}
\frac{1}{\mathbf{i}\hbar}\langle \underline{\Phi}^{{t}}_{\underline{H}_e}| [\hat{{\underline{X}}}^\alpha_e, \hat{{\underline{X}}}^\beta_e] |\underline{\Phi}^{{t}}_{\underline{H}_e}\rangle=\{{\underline{X}}^\alpha_e, {\underline{X}}^\beta_e\}+\mathcal{O}(t),
\end{equation}
where $\underline{H}_e=\{\underline{H}_e^\alpha\}$, $\underline{H}_e^\alpha=\underline{h}^\alpha_ee^{-\underline{Y}^\alpha_e} =e^{\mathbf{i}(\underline{\phi}^\alpha_e+\mathbf{i}\underline{Y}^\alpha_e)}$, and $({\underline{\tilde{h}}}_{e}, {\underline{X}}^\alpha_e)$ are defined by Eq.\eqref{repa} based on $(\underline{h}^\alpha_e,\underline{Y}^\alpha_e)$. One can conclude that the quantum algebras among $(\hat{\underline{\tilde{h}}}_{e}, \hat{{\underline{X}}}^\beta_e)$ acting in the quantum space $\underline{\mathcal{H}}$ spanned by the coherent states $\underline{\Phi}^{{t}}_{\underline{H}_e}(\underline{h}_e)$ gives a quantum representation of the Poisson algebras \eqref{hp33} among $({\underline{\tilde{h}}}_{e}, {\underline{X}}^\beta_e)$. Especially, this representation  endows with the interpretation of the all dimensional weak coupling LQG to $U(1)^{\frac{D(D+1)}{2}}$ loop quantum theory in the quantum subspace $\underline{\mathcal{H}}_w^c\subset \underline{\mathcal{H}}$ spanned by the coherent states $\underline{\Phi}^{{t}}_{\underline{H}_e}(\underline{h}_e)$ labelled by $\underline{H}_e$ with small $\underline{\phi}^\alpha_e$, since the Poisson algebras \eqref{hp33} among $({\underline{\tilde{h}}}_{e}, {\underline{X}}^\beta_e)$ coincide with those \eqref{hp220} among $({h}_{e}, {X}^\beta_e)$ in weak coupling limit given by small $\underline{\phi}^\alpha_e=\phi_e^\alpha$. From now on, we will restrict our discussion in the space $\underline{\mathcal{H}}_w^c$ and refer to the $U(1)^{\frac{D(D+1)}{2}}$ loop quantum theory as the weak coupling $U(1)^{\frac{D(D+1)}{2}}$ LQG.

%Now, we can claim that the $U(1)^{\frac{D(D+1)}{2}}$ loop quantum theory reveals the key quantum characters of $SO(D+1)$ LQG in the weak coupling limit. By this we mean that, in the weak coupling limit given by small $\underline{\phi}^\alpha_e=\phi_e^\alpha$, the basic symplectic structure of $SO(D+1)$ holonomy-flux phase space can be reproduced in $U(1)^{\frac{D(D+1)}{2}}$ holonomy-flux phase space based on the re-parametrization Eq.\eqref{repa}, so that the  $U(1)^{\frac{D(D+1)}{2}}$ loop quantum theory gives an identical description of weak coupling LQG  as the $SO(D+1)$  LQG for small $\underline{\phi}^\alpha_e=\phi_e^\alpha$.

The other operators in the weak coupling $U(1)^{\frac{D(D+1)}{2}}$ LQG can be established based on the re-constructed $SO(D +1)$ holonomy-flux operators. Generally, for an operator $\hat{O}=\hat{O}(\hat{h}_{e}, \hat{X}^\alpha_e)$ in the $SO(D+1)$ LQG, we can replace the $SO(D +1)$ holonomy-flux operators $(\hat{h}_{e}, \hat{X}^\alpha_e)$ in the expression $\hat{O}=\hat{O}(\hat{h}_{e}, \hat{X}^\alpha_e)$ by the re-constructed $SO(D +1)$ holonomy-flux operators $(\hat{\underline{\tilde{h}}}_{e}, \hat{\underline{X}}^\alpha_e)$ of the weak coupling $U(1)^{\frac{D(D+1)}{2}}$ LQG as
\begin{equation}\label{operatormap}
\hat{h}_{e}[A]\leftrightarrow\hat{\underline{\tilde{h}}}_{e}[{A}],\quad
\hat{X}^\alpha_e\leftrightarrow\hat{\underline{X}}^\alpha_e\quad\hat{\tilde{X}}^\alpha_e \leftrightarrow\hat{\tilde{\underline{X}}}^\alpha_e,
\end{equation}
to construct the corresponding operator $\underline{\hat{O}}=\hat{O}(\hat{\underline{\tilde{h}}}_{e}, \hat{\underline{X}}^\alpha_e)$ in the weak coupling $U(1)^{\frac{D(D+1)}{2}}$ LQG.
For instance,  one can replace $\hat{X}^\alpha_e$ and $\hat{\tilde{X}}^\alpha_e$ by $\hat{\underline{X}}^\alpha_e$ and $\hat{\tilde{\underline{X}}}^\alpha_e$ respectively in the definition \eqref{Vdef} of $\hat{V}_R$ in $SO(D+1)$ LQG, to establish the corresponding volume operator $\hat{\underline{V}}_R$ in the weak coupling $U(1)^{\frac{D(D+1)}{2}}$ LQG.

\section{Constraints in the weak coupling $U(1)^{\frac{D(D+1)}{2}}$ LQG}
Recall that the $U(1)^{\frac{D(D+1)}{2}}$ loop quantum theory only takes the interpretation of the all dimensional weak coupling LQG in the phase space region where the $U(1)^{\frac{D(D+1)}{2}}$ holonomy tends to identity. Thus, the physical consideration of the $U(1)^{\frac{D(D+1)}{2}}$ LQG should be restrict to the space $\underline{\mathcal{H}}_w^c$ composed by the quantum states whose wave functions are sharply peaked at the phase space region where the $U(1)^{\frac{D(D+1)}{2}}$ holonomy tends to identity. Nevertheless, we still need to solve the constraints in this theory to ensure that the quantum state takes correct physical degrees of freedom.
\subsection{The kinematic constraints}
Let us first consider the imposition of Gaussian and simplicity constraints in the weak coupling $U(1)^{\frac{D(D+1)}{2}}$ LQG.
The discrete version of the Gaussian constraint in $SO(D+1)$ LQG reads
\begin{equation}
\hat{G}_v^{\alpha}=\sum_{e,s(e)=v}\hat{X}^\alpha_e+\sum_{e,t(e)=v}\hat{\tilde{X}}^\alpha_e=0.
\end{equation}
Then, the corresponding discrete ``Gaussian constraint'' in the weak coupling $U(1)^{\frac{D(D+1)}{2}}$ LQG can be given directly as
\begin{equation}
\hat{\underline{G}}_v^{\alpha} =\sum_{e,s(e)=v}\hat{\underline{X}}^\alpha_e+\sum_{e,t(e)=v}\hat{\tilde{\underline{X}}}^\alpha_e=0.
\end{equation}
Note that this ``Gaussian constraint'' does not generate the $U(1)^{\frac{D(D+1)}{2}}$ gauge transformations. In fact, it is just the closure condition for the $D$-polytopes described by its oriented $(D-1)$-areas \cite{Long:2020agv,PhysRevD.103.086016}.
Similarly, the quantum edge-simplicity and vertex-simplicity constraints  in the weak coupling $U(1)^{\frac{D(D+1)}{2}}$ LQG can be given as
\begin{equation}\label{weaksimedge}
\underline{S}_e^{IJKL}\equiv \hat{\underline{X}}^{[IJ}_e \hat{\underline{X}}^{KL]}_e\approx0, \ \forall e\in \Gamma
\end{equation}
and
\begin{equation}\label{weaksimvertex}
 \underline{S}_{v,e,e'}^{IJKL}\equiv \hat{\underline{X}}^{[IJ}_e \hat{\underline{X}}^{KL]}_{e'}\approx0,\ \forall e,e'\in \Gamma, s(e)=s(e')=v
\end{equation}
respectively, wherein $\hat{\underline{X}}^{IJ}_e:=\hat{\underline{X}}^\alpha_e\tau_\alpha^{IJ}$.

The imposition of the Gaussian and simplicity constraints in weak coupling $U(1)^{\frac{D(D+1)}{2}}$ LQG is different with that in $SO(D+1)$ LQG. Recall that the Gaussian and edge-simplicity constraints in $SO(D+1)$ LQG eliminate the degrees of freedom by solving the constraint equations and taking the averaging with respect to the corresponding gauge transformations, while the vertex simplicity constraint in $SO(D+1)$ LQG eliminate the degrees of freedom by solving the constraint equation. However, one should notice that the Gaussian and simplicity constraints in the weak coupling $U(1)^{\frac{D(D+1)}{2}}$ LQG only generate correct gauge transformations in the phase space region where the $U(1)^{\frac{D(D+1)}{2}}$ holonomy tends to identity. Thus, it is not valid to eliminate the gauge degrees of freedom by taking the averaging in the weak coupling $U(1)^{\frac{D(D+1)}{2}}$ LQG.

In order to eliminate the degrees of freedom constrained by the Gaussian and simplicity constraint in the weak coupling $U(1)^{\frac{D(D+1)}{2}}$ LQG, let us consider another strategy, that is, one can weakly solve the corresponding constraint equations and then take the gauge fixing. Notice that the heat-kernel coherent states of $U(1)^{\frac{D(D+1)}{2}}$ have well-behaved peakedness property, one can also proceed this strategy based on this coherent states at the semi-classical level. Let us now consider it in details.

The weak imposition of the quantum Gaussian and simplicity constraints based on the heat-kernel coherent states of $U(1)^{\frac{D(D+1)}{2}}$ can be given as
\begin{equation}\label{weakGauss}
\langle \underline{\Phi}^{{t}}_{\Gamma,\underline{H}'}|\hat{\underline{G}}_v^{\alpha} |\underline{\Phi}^{{t}}_{\Gamma,\underline{H}}\rangle =\sum_{e,s(e)=v}\underline{X}^\alpha_e(\underline{H}_e)+\sum_{e,t(e)=v}\tilde{\underline{X}}^\alpha_e(\underline{H}_e)+\mathcal{O}(t)=0
\end{equation}
and
\begin{equation}\label{weakedgesim}
\langle \underline{\Phi}^{{t}}_{\Gamma,\underline{H}'}|\underline{S}_e^{IJKL} |\underline{\Phi}^{{t}}_{\Gamma,\underline{H}}\rangle =\underline{X}^\alpha_e(\underline{H}_e)\underline{X}^\beta_e(\underline{H}_e)\tau_\alpha^{[IJ}\tau_\beta^{KL]}+\mathcal{O}(t)=0,
\end{equation}
\begin{equation}\label{weakversim}
\langle \underline{\Phi}^{{t}}_{\Gamma,\underline{H}'}|\underline{S}_{v,e,e'}^{IJKL} |\underline{\Phi}^{{t}}_{\Gamma,\underline{H}}\rangle =\underline{X}^\alpha_e(\underline{H}_e)\underline{X}^\beta_{e'}(\underline{H}_{e'})\tau_\alpha^{[IJ}\tau_\beta^{KL]}+\mathcal{O}(t)=0,
\end{equation}
where $e,e'\in \Gamma, s(e)=s(e')=v$, $\underline{H}^\alpha_e=\underline{h}^\alpha_ee^{-\underline{Y}^\alpha_e}$, and
 \begin{equation} \underline{X}^\alpha_e(\underline{H}_e)\tau_\alpha=\underline{\tilde{h}}_{e}[{A}/2]\tau_\alpha \underline{\tilde{h}}^{-1}_{e}[{A}/2]{\underline{Y}}^\alpha_e,
 \end{equation}
\begin{equation}
\tilde{\underline{X}}^\alpha_e(\underline{H}_e)\tau_\alpha=-\underline{\tilde{h}}^{-1}_{e}[{A}/2]\tau_\alpha \underline{\tilde{h}}_{e}[{A}/2]{\underline{Y}}^\alpha_e
\end{equation}
with $\underline{\tilde{h}}_{e}[{A}/2]:=\exp\left(\frac{\sum_{\alpha}\left(\underline{h}_{e}^\alpha[{A}] -(\underline{h}_{e}^\alpha[{A}])^{-1}\right)\tau_i}{4\mathbf{i}}\right)$. The labels $\underline{H}$ of the solution states $\underline{\Phi}^{{t}}_{\Gamma,\underline{H}}$ of Eqs.\eqref{weakedgesim} and \eqref{weakversim} satisfy the conditions
\begin{equation}\label{solusim}
\underline{X}^\alpha_e(\underline{H}_e)\tau_\alpha^{IJ}=\underline{N}_v^{[I}\underline{X}^{J]}_e(\underline{H}_e),\quad \tilde{\underline{X}}^\alpha_{e'}(\underline{H}_e')\tau_\alpha^{IJ}=\underline{N}_v^{[I} \tilde{\underline{X}}^{J]}_{e'}(\underline{H}_e),\quad \forall \ e,e'\in \Gamma, s(e)=t(e')=v
\end{equation}
at leading order of $t$, where $\underline{N}_v^{I}$ is an unit vector at $v$. With the condition \eqref{solusim} being satisfied,  $\underline{N}_v^{I}$, $\underline{X}^{J}_e$ and $\tilde{\underline{X}}^{J}_{e}$ can be determined by $\underline{H}$. One can further solve Eq.\eqref{weakGauss},  which leads that the labels $\underline{H}$ of the weak solution states $\underline{\Phi}^{{t}}_{\Gamma,\underline{H}}$  of the quantum Gaussian and simplicity constraints satisfy
\begin{equation}\label{solugausssim2}
\sum_{e,s(e)=v}\underline{X}^I_e(\underline{H})+\sum_{e,t(e)=v}\tilde{\underline{X}}^I _e(\underline{H})=0,\quad v\in\Gamma
\end{equation}
and the condition \eqref{solusim}
at leading order of $t$.

Though the Gaussian and simplicity constraint equations are solved weakly with the  condition \eqref{solusim} and \eqref{solugausssim2} being satisfied, the gauge reduction has not been complete yet. Since arbitrary two phase space points related by the gauge transformations take the same physical interpretation, one can reduce the gauge degrees of freedom by identifying those states $\underline{\Phi}^{{t}}_{\Gamma,\underline{H}}$, whose labels $\underline{H}$ satisfying the condition \eqref{solusim} and \eqref{solugausssim2} are related by the gauge transformations induced by Gaussian and simplicity constraints. In fact, in order to proceed specific analysis, one always need to take an arbitrary but fixed gauge to choose a gauge fixing state in each set of the identified states. Besides, the operators corresponding physical observables must be gauge invariant, which can be constructed by generalizing the gauge invariant operators in $SO(D+1)$ LQG to the weak coupling $U(1)^{\frac{D(D+1)}{2}}$ LQG based on the relation  \eqref{operatormap}.

Usually, the gauge invariant operator with respect to simplicity constraint in $SO(D+1)$ LQG should be constructed based on the gauge averaging scheme, since the edge-simplicity constraint  in $SO(D+1)$ LQG are imposed strongly. We have shown in section \ref{scalaroperator} that the scalar constraint operator \eqref{scalarop} in $SO(D+1)$ LQG  is constructed based on gauge fixing scheme erroneously, which leads that it does have correct geometric interpretation. However, since the simplicity constraint in the weak coupling $U(1)^{\frac{D(D+1)}{2}}$ LQG is solved weakly and the corresponding gauge degrees of freedom are eliminated by gauge fixing, the gauge invariant operator with respect to simplicity constraint in the weak coupling $U(1)^{\frac{D(D+1)}{2}}$ LQG should be constructed based on the gauge fixing scheme. As we will see in next subsection, the scalar constraint operator \eqref{scalarop} constructed based on gauge fixing scheme can be generalized to the weak coupling $U(1)^{\frac{D(D+1)}{2}}$ LQG with correct geometric interpretation.

 \subsection{The ADM constraints}
 Let us consider the treatment of diffeomorphism and scalar constraints in the weak coupling $U(1)^{\frac{D(D+1)}{2}}$ LQG in this subsection. In $SO(D+1)$ LQG, the degrees of freedom constrained by diffeomorphism constraint are eliminated by taking the averaging over all of the diffeomorphism transformation on the $D$-manifold $\sigma$. This treatment can be generalized to the weak coupling $U(1)^{\frac{D(D+1)}{2}}$ LQG directly, since the spin-network states in $SO(D+1)$ LQG and charge-network states in the weak coupling $U(1)^{\frac{D(D+1)}{2}}$ LQG are established on graphes, and their diffeomorphism transformation only involve the deformation and translation of these graphs.

 The construction of the scalar constraint operator in the weak coupling $U(1)^{\frac{D(D+1)}{2}}$ LQG will be different with the one in $SO(D+1)$ LQG a lot. Recall that the reduced $SO(D+1)$ holonomy in $SO(D+1)$ LQG can not capture the degrees of freedom of spatial intrinsic curvature, so that it can not been used to construct the scalar constraint operator by the standard strategy. Nevertheless, the simplicity constraint in the weak coupling $U(1)^{\frac{D(D+1)}{2}}$ LQG is treated in a different strategy, by this it means that, we only weakly solve the simplicity constraint equation, but do not take the averaging with respect to the gauge transformation or  make the gauge fixing. Hence, as one kind of smearing versions of connection, the re-constructed $SO(D+1)$ holonomy in the weak coupling $U(1)^{\frac{D(D+1)}{2}}$ LQG has the same geometric interpretation as the original $SO(D+1)$ holonomy in $SO(D+1)$ LQG, which means that it captures the degrees of freedom of both extrinsic and intrinsic curvature, as well as the gauge degrees of freedom with respect to the simplicity constraint. Now, let us recall the scalar constraint operator in $SO(D+1)$ LQG and consider the construction of the scalar constraint operator in the weak coupling $U(1)^{\frac{D(D+1)}{2}}$ LQG.

Notice that Eq.\eqref{scalarop} is not a correct scalar constraint operator  in $SO(D+1)$ LQG, since the holonomy operator in its expression takes a different geometric interpretation from the classical holonomy. Nevertheless, this problem can be avoided in the weak coupling $U(1)^{\frac{D(D+1)}{2}}$ LQG. Notice that we only solve the simplicity constraint equations weakly and no degrees of freedom in the re-constructed $SO(D+1)$ holonomy are eliminated. Thus, the re-constructed $SO(D+1)$ holonomy captures the geometric degrees of freedom properly. One can re-construct the scalar constraint operator $\underline{\hat{C}}[N]$ in the weak coupling $U(1)^{\frac{D(D+1)}{2}}$ LQG by replacing $\hat{h}_e$, ${\hat{X}}^\alpha_e$ and $\hat{\tilde{X}}^\alpha_e$ by  $\hat{\underline{\tilde{h}}}_e$, ${\hat{\underline{X}}}^\alpha_e$ and $\hat{\tilde{\underline{X}}}^\alpha_e$ respectively in the definition \eqref{scalarop} of the scalar constraint operator in $SO(D+1)$ LQG, which leads to
\begin{equation}\label{scalarweak}
\underline{\hat{C}}[N]=\underline{\hat{C}}_{\text{E}}[N]+\underline{\hat{C}}_{\text{L}}[N] +\underline{\hat{C}}_{\text{D}}[N],
\end{equation}
 where $\underline{\hat{C}}_{\text{E}}[N]$, $\underline{\hat{C}}_{\text{L}}[N] $ and $\underline{\hat{C}}_{\text{D}}[N]$ are given by substituting $(\hat{h}_e,\hat{X}^\alpha_e, \hat{\tilde{X}}^\alpha_e)$ with $(\hat{\underline{\tilde{h}}}_e, \hat{\underline{X}}^\alpha_e,\hat{\tilde{\underline{X}}}^\alpha_e)$ respectively in the expressions \eqref{CECL} and \eqref{CD} of ${\hat{C}}_{\text{E}}[N]$, ${\hat{C}}_{\text{L}}[N] $ and ${\hat{C}}_{\text{D}}[N]$.
 \section{Conclusion and Outlook}
The weak coupling loop quantum theory with Abelian gauge group provides us a new perspective to study the weak coupling properties of LQG. In this paper, the loop quantization of the connection formulation of $(1+D)$-dimensional GR is given based on the $U(1)^{\frac{D(D+1)}{2}}$ holonomy-flux algebra. It is shown that the $SO(D+1)$ holonomy-flux Poisson algebra can be re-produced based on the $U(1)^{\frac{D(D+1)}{2}}$ holonomy-flux Poisson algebra in the weak coupling limit, with the $SO(D+1)$ holonomy-flux phase space being parametrized by  $U(1)^{\frac{D(D+1)}{2}}$ holonomy-flux. Thus, it is reasonable to claim that the $U(1)^{\frac{D(D+1)}{2}}$ loop quantum theory gives another kind of loop representation of the $SO(D+1)$ holonomy-flux Poisson algebra in weak coupling limit. Then, by generalizing the constraint operators in  the $SO(D+1)$ LQG to the $U(1)^{\frac{D(D+1)}{2}}$ loop quantum theory, the weak coupling $U(1)^{\frac{D(D+1)}{2}}$ LQG can be established with the Hilbert space being spanned by the  $U(1)^{\frac{D(D+1)}{2}}$ coherent states peaked at the weak coupling region of the phase space.

 It has been verified that the classical scalar constraint in connection formulation of $(1+D)$-dimensional GR can not be used to construct the scalar constraint operator in the $SO(D+1)$ LQG, since the gauge reduction with respect to the simplicity constraint is proceeded by using the gauge fixing method in classical connection theory, while it is proceeded by using the gauge transformation averaging method in the $SO(D+1)$ LQG.
 Different with the $SO(D+1)$ LQG, the gauge reduction with respect to the simplicity constraint is proceeded by using the gauge fixing method in the weak coupling $U(1)^{\frac{D(D+1)}{2}}$ LQG. More explicitly, the Gaussian  and simplicity constraints are imposed weakly based on the $U(1)^{\frac{D(D+1)}{2}}$ heat-kernel coherent states, and the solution states are given by those coherent states satisfying conditions \eqref{solusim} and \eqref{solugausssim2}. Thus, the scalar constraint operator \eqref{scalarweak} in the weak coupling $U(1)^{\frac{D(D+1)}{2}}$ LQG is constructed by regularizing and quantizing the classical scalar constraint in connection formulation of $(1+D)$-dimensional GR.

Several interesting points of the weak coupling $U(1)^{\frac{D(D+1)}{2}}$ LQG deserve further investigation. First, the $U(1)^{\frac{D(D+1)}{2}}$ heat-kernel coherent state in the weak coupling $U(1)^{\frac{D(D+1)}{2}}$ LQG and the twisted geometry coherent states in $SO(D+1)$ LQG are both expected to provide some kind of semi-classical description of the $(1+D)$-dimensional spacetime geometry  \cite{Long:2021xjm,Long:2021lmd,Long:2022cex}. Hence, it is worth to compare the properties of these two kinds of coherent states. Especially, since the twisted geometry coherent states in $SO(D+1)$ LQG strongly vanish the edge-simplicity constraint while the $U(1)^{\frac{D(D+1)}{2}}$ heat-kernel coherent state weakly solve the edge-simplicity constraint, it is interesting to explore how to capture the physical degrees of freedom correctly by defining the operators corresponding to kinds of physical observables in these two theories.
 Second, one can consider the effective dynamics of the weak coupling $U(1)^{\frac{D(D+1)}{2}}$ LQG based on the coherent states and the scalar constraint operator. Since the gauge degrees of freedom with respect to Gaussian and simplicity constraint are eliminated by gauge fixing, it is necessary to verify that the effective dynamics of the weak coupling $U(1)^{\frac{D(D+1)}{2}}$ LQG  are independent to the choices the gauge fixing. Third,  it has been shown that the Hamiltonians of the matter fields can be defined
in the weak coupling $U(1)^3$ LQG coupled with matters in $(1+3)$-dimensional spacetime \cite{Sahlmann:2002qj}. One can generalize this study to the weak coupling $U(1)^{\frac{D(D+1)}{2}}$ LQG, to construct the Hamiltonian operator for the the weak coupling $U(1)^{\frac{D(D+1)}{2}}$ LQG coupled with matters and
study its dynamics. Usually, in the case
where quantum field theory (QFT) on curved spacetimes is valid, the spacetime curvature is not too
big. Then, one can further understand this weak field situation by assuming all of the holonomies
in all dimensional LQG approach to identity such that the weak coupling condition is satisfied. Moreover, only the
effective semiclassical geometry and its dynamics is concerned as the background of QFT. Hence,
the weak coupling $U(1)^{\frac{D(D+1)}{2}}$ LQG  with much simpler revelent calculations is a good alternative of the $SO(D+1)$ LQG for exploring whether QFT on curved $(1+D)$-dimensional spacetimes could be obtained as certain semiclassical limit of all dimensional LQG. Especially, the Fermions coupling to $SO(D+1)$ LQG involves the non-simple representations of $SO(D+1)$, which contradicts to the strong imposition of the edge-simplicity constraint. The weak coupling $U(1)^{\frac{D(D+1)}{2}}$ LQG may provide a new perspective to deal with this issue, since the edge-simplicity constraint is imposed weakly in this theory.

\section*{Acknowledgments}
This work is supported by the project funded by China
Postdoctoral Science Foundation  with Grant No. 2021M691072, and the National Natural Science Foundation of China (NSFC) with Grants No. 12047519, No. 11875006 and No. 11961131013.

\bibliographystyle{unsrt}

\bibliography{ref}

\begin{thebibliography}{10}

\bibitem{Ashtekar2012Background}
Abhay Ashtekar and Jerzy Lewandowski.
\newblock Background independent quantum gravity: a status report.
\newblock {\em Classical and Quantum Gravity}, 21(15):R53--R152, 2012.

\bibitem{RovelliBook2}
Carlo Rovelli and Francesca Vidotto.
\newblock {\em {Covariant Loop Quantum Gravity: An Elementary Introduction to
  Quantum Gravity and Spinfoam Theory}}.
\newblock Cambridge University Press, 2014.

\bibitem{Han2005FUNDAMENTAL}
Muxin Han, M.~A. Yongge, and Weiming Huang.
\newblock Fundamental structure of loop quantum gravity.
\newblock {\em International Journal of Modern Physics D}, 16(09):1397--1474,
  2005.

\bibitem{thiemann2007modern}
Thomas Thiemann.
\newblock {\em Modern canonical quantum general relativity}.
\newblock Cambridge University Press, 2007.

\bibitem{rovelli2007quantum}
Carlo Rovelli.
\newblock {\em Quantum gravity}.
\newblock Cambridge university press, 2007.

\bibitem{Smolin_1992}
L~Smolin.
\newblock The $\textsc{G}_{\textsc{n}\text{ewton}}$ to 0 limit of euclidean
  quantum gravity.
\newblock {\em Classical and Quantum Gravity}, 9(4):883--893, apr 1992.

\bibitem{PhysRevD.87.044039}
Casey Tomlin and Madhavan Varadarajan.
\newblock Towards an anomaly-free quantum dynamics for a weak coupling limit of
  euclidean gravity.
\newblock {\em Phys. Rev. D}, 87:044039, Feb 2013.

\bibitem{Lewandowski:2016lby}
Jerzy Lewandowski and Chun-Yen Lin.
\newblock {Exploring the Tomlin-Varadarajan quantum constraints in $U(1)^3$
  loop quantum gravity: solutions and the Minkowski theorem}.
\newblock {\em Phys. Rev. D}, 95(6):064032, 2017.

\bibitem{Sahlmann:2002qj}
Hanno Sahlmann and Thomas Thiemann.
\newblock {Towards the QFT on curved space-time limit of QGR. 1. A General
  scheme}.
\newblock {\em Class. Quant. Grav.}, 23:867--908, 2006.

\bibitem{Sahlmann:2002qk}
Hanno Sahlmann and Thomas Thiemann.
\newblock {Towards the QFT on curved space-time limit of QGR. 2. A Concrete
  implementation}.
\newblock {\em Class. Quant. Grav.}, 23:909--954, 2006.

\bibitem{Long:2021izw}
Gaoping Long and Yongge Ma.
\newblock {Effective dynamics of weak coupling loop quantum gravity}.
\newblock {\em Phys. Rev. D}, 105(4):044043, 2022.

\bibitem{Bodendorfer:Ha}
Norbert Bodendorfer, Thomas Thiemann, and Andreas Thurn.
\newblock New variables for classical and quantum gravity in all dimensions: I.
  \textsc{H}amiltonian analysis.
\newblock {\em Classical and Quantum Gravity}, 30(4):045001, 2013.

\bibitem{Bodendorfer:La}
Norbert Bodendorfer, Thomas Thiemann, and Andreas Thurn.
\newblock New variables for classical and quantum gravity in all dimensions:
  \textsc{II}. \textsc{L}agrangian analysis.
\newblock {\em Classical and Quantum Gravity}, 30(4):045002, 2013.

\bibitem{Bodendorfer:Qu}
Norbert Bodendorfer, Thomas Thiemann, and Andreas Thurn.
\newblock New variables for classical and quantum gravity in all dimensions:
  \textsc{III}. \textsc{Q}uantum theory.
\newblock {\em Classical and Quantum Gravity}, 30(4):045003, 2013.

\bibitem{Bodendorfer:SgI}
Norbert Bodendorfer, Thomas Thiemann, and Andreas Thurn.
\newblock Towards loop quantum supergravity (lqsg): I.
  \textsc{R}arita--\textsc{S}chwinger sector.
\newblock {\em Classical and Quantum Gravity}, 30(4):045006, 2013.

\bibitem{Bodendorfer:2011onthe}
Norbert Bodendorfer, Thomas Thiemann, and Andreas Thurn.
\newblock On the implementation of the canonical quantum simplicity constraint.
\newblock {\em Classical and Quantum Gravity}, 30(4):045005, 2013.

\bibitem{PhysRevD.103.086016}
Gaoping Long and Chun-Yen Lin.
\newblock Geometric parametrization of $so\mathbf{(}d+1\mathbf{)}$ phase space
  of all dimensional loop quantum gravity.
\newblock {\em Phys. Rev. D}, 103:086016, Apr 2021.

\bibitem{long2019coherent}
Gaoping Long, Chun-Yen Lin, and Yongge Ma.
\newblock Coherent intertwiner solution of simplicity constraint in all
  dimensional loop quantum gravity.
\newblock {\em Physical Review D}, 100(6):064065, 2019.

\bibitem{Long:2022thb}
Gaoping Long and Xiangdong Zhang.
\newblock {On the gauge reduction with respect to simplicity constraint in all
  dimensional loop quantum gravity}.
\newblock 9 2022.

\bibitem{Long:2020euh}
Gaoping Long and Norbert Bodendorfer.
\newblock {Perelomov-type coherent states of SO($D+1$) in all-dimensional loop
  quantum gravity}.
\newblock {\em Phys. Rev. D}, 102(12):126004, 2020.

\bibitem{Calcinari_2020}
Andrea Calcinari, Laurent Freidel, Etera Livine, and Simone Speziale.
\newblock Twisted geometries coherent states for loop quantum gravity.
\newblock {\em Classical and Quantum Gravity}, 38(2):025004, Dec 2020.

\bibitem{Long:2021xjm}
Gaoping Long, Cong Zhang, and Xiangdong Zhang.
\newblock {Superposition type coherent states in all dimensional loop quantum
  gravity}.
\newblock {\em Phys. Rev. D}, 104(4):046014, 2021.

\bibitem{Long:2021lmd}
Gaoping Long, Xiangdong Zhang, and Cong Zhang.
\newblock {Twisted geometry coherent states in all dimensional loop quantum
  gravity: Construction and peakedness properties}.
\newblock {\em Phys. Rev. D}, 105(6):066021, 2022.

\bibitem{Long:2022cex}
Gaoping Long.
\newblock {Twisted geometry coherent states in all dimensional loop quantum
  gravity: II. Ehrenfest Property}.
\newblock 4 2022.

\bibitem{long2020operators}
Gaoping Long and Yongge Ma.
\newblock {General geometric operators in all dimensional loop quantum
  gravity}.
\newblock {\em Phys. Rev. D}, 101(8):084032, 2020.

\bibitem{Long:2020agv}
Gaoping Long and Yongge Ma.
\newblock {Polytopes in all dimensional loop quantum gravity}.
\newblock {\em Eur. Phys. J. C}, 82(1):41, 2022.

\bibitem{Zhang:2015bxa}
Xiangdong Zhang.
\newblock {Higher dimensional Loop Quantum Cosmology}.
\newblock {\em Eur. Phys. J. C}, 76(7):395, 2016.

\end{thebibliography}

%\appendix
%\section{}

\end{document}